\documentclass{article}
\usepackage{graphicx} 
\usepackage{subcaption}
\usepackage{amsmath}
\usepackage{float}
\usepackage{booktabs}
\usepackage{geometry}
\usepackage{hyperref}
\usepackage[round]{natbib}
\usepackage{authblk}
\title{An update to ECMWF's machine-learned weather forecast model AIFS}

\author[1]{Gabriel Moldovan$^{*}$}
\author[1]{Ewan Pinnington$^{*}$}
\author[1]{Ana Prieto Nemesio\thanks{equal contribution}}
\author[1]{Simon Lang}
\author[1]{Zied Ben Bouallègue}
\author[1]{Jesper Dramsch}
\author[1]{Mihai Alexe}
\author[1]{Mario Santa Cruz}
\author[1]{Sara Hahner}
\author[1]{Harrison Cook}
\author[1]{Helen Theissen}
\author[1]{Mariana Clare}
\author[1]{Cathal O'Brien}
\author[1]{Jan Polster}
\author[1]{Linus Magnusson}
\author[1]{Gert Mertes}
\author[1]{Florian Pinault}
\author[1]{Baudouin Raoult}
\author[1]{Patricia de Rosnay}
\author[1]{Richard Forbes}
\author[1]{Matthew Chantry}

\affil[1]{European Centre for Medium-Range Weather Forecasts}

\date{March 2025}

\begin{document}

\maketitle

\abstract

We present an update to ECMWF's machine-learned weather forecasting model AIFS Single with several key improvements. The model now incorporates physical consistency constraints through bounding layers, an updated training schedule, and an expanded set of variables. The physical constraints substantially improve precipitation forecasts and the new variables show a high level of skill. Upper-air headline scores also show improvement over the previous AIFS version. The AIFS has been fully operational at ECMWF since the 25th of February 2025.

\section{Introduction}
Machine-learned weather forecast models have started to rival or outperform physics-based numerical weather prediction (NWP) models in recent years \citep{pathak2022fourcastnet, keisler2022forecasting, lam2022graphcast, Chen2023, bi2023accurate, lang2024aifs}. For both training and forecasting, these machine-learned forecast models mostly depend on the Copernicus ERA5 reanalysis dataset produced by ECMWF \citep{hersbach2020era5} and operational analysis by ECMWF's physics-based integrated forecasting system (IFS).

ECMWF has developed the artificial intelligence forecasting system (AIFS) \citep{lang2024aifs}, its own machine-learned forecast model. After a successful pre-operational test phase running four times daily since October 2023, with forecasts publicly available under ECMWF's open data policy, AIFS has now transitioned to operational status. The first operational version, AIFS 1.0.0 replacing AIFS 0.2.1, was implemented on 25 February 2025. The current operational version, AIFS 1.1.0 described here, was released on 27 August 2025 to correct a precipitation forecast issue in the initial version. The model is trained with a mean-squared error (MSE) loss function and is referred to as AIFS Single, to distinguish it from the probabilistically trained version, the AIFS ENS \citep{lang2024aifs-crps}. 

Although such MSE-trained forecast models have been shown to smooth forecast fields at longer lead times to avoid the double-penalty of incorrectly positioned weather phenomena \citep{lam2022graphcast, benbouallegue2023rise, lang2024aifs, bonavita2024}, they still display physically robust characteristics \citep{hakim2024dynamical} and are able to make useful predictions of extreme events \citep{benbouallegue2023rise}. The cheaper training costs associated with MSE-trained models (compared to probabilistically trained models) make them attractive for prototyping new features and model components.

To date, most machine-learned weather forecast models only include a limited subset of forecast variables available from current NWP systems. Here, we include for the first time in the AIFS soil moisture, soil temperature and runoff together with energy sector variables such as cloud cover, 100 metre winds and solar radiation. The choice of additional variables has been guided by utility to users and with considerations of future applications of the model, alongside pragmatic considerations on data availability and readiness. Surface solar radiation and 100-metre wind speeds have been included, important for renewable energy sectors. We added an initial characterization of the land surface with prognostic soil moisture and soil temperature, important for drought forecasting. We also include snowfall, improving the representation of distinct precipitation types in the model. Finally, we have added run-off as a diagnostic model output, pushing towards a hydrological component for the AIFS.

Despite their ability to produce skilful forecasts, machine-learned forecast models are prone to producing outputs that violate known physical relationships and limits (e.g., negative precipitation or mass imbalances). In current applications, including the pre-operational version of AIFS, post-processing of forecasts is commonly applied to remove such physical inconsistencies. Instead, we propose an additional final layer of activation functions that bound certain variables within physically meaningful limits and enforce physical constraints between related quantities. This simplifies the learning task by constraining the model output space to physically plausible regimes. This bounding strategy also proves particularly beneficial for variables with non-Gaussian distributions, such as precipitation, where the model must effectively distinguish between rain and no-rain states. The bounding layer effectively maps negative outputs to no-rain, eliminating the need for the model to explicitly learn to predict zero-precipitation values.

In this paper we begin by outlining the training setup of the model and how this differs from the previous AIFS version. Then we motivate and describe the new bounding strategy to make the model forecast more physically consistent. We demonstrate the improved performance of the revised AIFS version via evaluation results and selected case studies. We conclude by summarizing main results and future work in the discussion and conclusions.

\section{Training}
The architecture of AIFS follows an encoder-processor-decoder design. Here, encoder and decoder are attention-based graph neural networks, and the processor is a transformer with a sliding window attention (see \cite{lang2024aifs} for details). 


The model operates on a reduced Gaussian grid, (N320, approximately 0.25° resolution). The processor (or hidden) grid is an O96 octahedral reduced Gaussian grid (\cite{Wedi2014}) with 40,320 grid points, approximately 1° resolution, and consists of 16 processor layers.

AIFS is trained to produce 6-hour forecasts $t_{+6\text{h}}$ using past and present atmospheric states at $t_{-6\text{h}}$ and $t_0$ (from ERA5 or ECMWF’s operational analyses at initialization, or from the model forecast itself). Longer lead times are produced auto-regressively by feeding the model’s predictions back as inputs, a process commonly referred to as rollout.

\subsection{Training Schedule}
The training is divided into two phases. The first is a pre-training phase, where the model learns to predict the atmospheric state 6 hours ahead ($t_{+6\text{h}}$) using ERA5 analysis at $t_{-6\text{h}}$ and $t_0$.
The second phase, rollout fine-tuning, continues from the pre-trained weights and trains the model to forecast auto-regressively up to 72 hours. Here, the model learns to forecast from its own predictions.
Unlike the previous AIFS version, where rollout fine-tuning was first performed using ERA5 and then followed by final fine-tuning on ECMWF operational analysis, we directly use operational analysis for the entire fine-tuning stage. This simplifies the training pipeline, reduces computational costs and results in better forecast performance. 


Pre-training is performed on ERA5 data covering the years 1979–2022 (compared to 1979–2020 in the previous AIFS version), using a cosine learning rate (LR) schedule, a batch size of 16, and a total of 260,000 training steps. The LR is linearly increased from 0 to $5 \times 10^{-4}$ during the first 1,000 steps, then annealed to a minimum of $3 \times 10^{-7}$. This is followed by rollout fine-tuning on ECMWF operational analysis from 2016 to 2022, also using a cosine LR schedule and batch size of 16, for approximately 7,900 steps (equivalent to one epoch per rollout step). The LR started at $1.28 \times 10^{-5}$ and is annealed to the same minimum value of $3 \times 10^{-7}$. The rollout length is initially set to 6 hours (1 step) and progressively increased by one step per epoch up to 72 hours (12 steps), following the approach of \cite{lam2022graphcast} and \cite{lang2024aifs}. We used the AdamW optimizer \citep{loshchilov2018decoupled} with $\beta$ coefficients of 0.9 and 0.95. Here, the rollout dataset is extended to eight years of operational IFS analysis (2016–2022), compared with only two years (2019–2020) in the previous AIFS version.


\subsection{Variables used in training}
\begin{figure}
    \centering
    \includegraphics[width=0.7\linewidth]{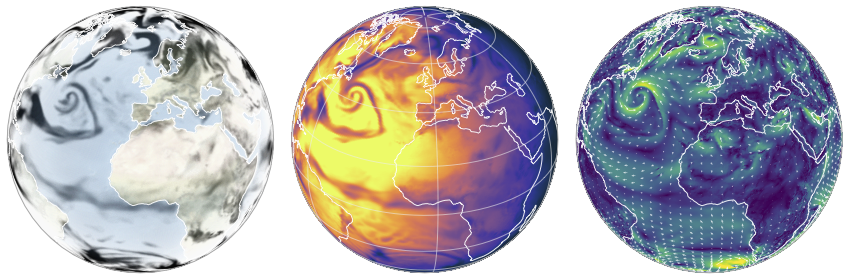}
\caption{A selection of new variables available from the revised AIFS Single forecasts: cloud cover (left), surface solar radiation (centre), and 100 m wind speed/direction (right). The consistency between these new variables is clear, with areas of higher cloud cover corresponding to lower solar radiation at the surface and consistent weather patterns for 100-metre winds.}
\label{fig:new_vars}
\end{figure}

 The variables used in the new AIFS version are listed in Table \ref{tab:aifs_io_variables}. As in AIFS 0.2.1, the upper atmosphere is represented by geopotential, horizontal wind components, specific humidity, and temperature at 13 pressure levels: 50, 100, 150, 200, 250, 300, 400, 500, 600, 700, 850, 925, and 1000 hPa. Newly introduced variables are marked with *. We have increased the characterization of the land surface in the model by including new prognostic variables of soil moisture at levels 1 and 2 (swvl1 and swvl2), and soil temperature at levels 1 and 2 (stl1 and stl2), important for drought monitoring and forecasting. A notion of hydrology has been included with runoff (ro), forecast as a diagnostic variable. A second set of variables, related to energy forecasting and clouds, adds real value to the model’s utility. These are forecast diagnostically and include the 100-metre wind components (100u and 100v), surface solar and thermal radiation (ssrd and strd), and cloud cover at various levels (tcc, hcc, mcc, lcc). Finally, snowfall (sf) has been added to complement the set of total precipitation–related variables. An illustration of a selection of these variables can be seen in the forecast presented in Figure \ref{fig:new_vars}, where the consistency between these new variables is clear, with areas of higher cloud cover corresponding to lower solar radiation at the surface and consistent weather patterns for 100-metre winds. These new variables are sourced from the ERA5 reanalysis and IFS operational data archive, in line with those used in the previous AIFS version (0.2.1).

The per variable normalization strategy used in AIFS is summarized in Table \ref{tab:aifs_io_variables}. Unless stated otherwise, data is normalized to zero mean and unit variance (z-score normalization). For some bounded output variables (see Section \ref{sec:bounding}), only standard deviation normalization is applied to avoid shifting of the absolute zero in the normalized space. The loss function is unchanged from the previous AIFS version. Table \ref{tab:aifs_io_variables} shows the loss scaling factors we use in the revised AIFS version. Scaling factors were chosen empirically to ensure that all prognostic variables contribute approximately equally to the loss function, with the exception of vertical velocities and soil moisture, deliberately down-weighted. Furthermore, the loss weights decrease linearly with height, so that upper atmospheric levels contribute less to the total loss. The pressure level weights are calculated following $w = \max(\text{pressure level}/1000, 0.2)$, like in the AIFS-ENS \citep{lang2024aifs-crps}. A minimum weight of 0.2 is imposed in the revised version to avoid assigning excessively low values in the stratosphere. 

AIFS is trained using data parallelism with a batch size of 16, while each model instance is distributed across four GPUs within a single node \citep{lang2024aifs}. Training was conducted on the European supercomputer Leonardo (EuroHPC), hosted and managed by Cineca, on 64GB A100 GPUs. Mixed-precision training is used (\cite{micikevicius2018mixedprecisiontraining}), and the full process takes approximately three days. A 10-day forecast can be produced in about 2 minutes and 30 seconds on a single A100 GPU, including data input and output.

\begin{table}[htbp]
\centering
\small
\renewcommand{\arraystretch}{1} 
\begin{tabular}{|p{4.2cm}|p{1.5cm}|p{1.7cm}|p{1.5cm}|p{2.2cm}|p{1.25cm}|}
\hline
\textbf{Variable name} & \textbf{Short name} & \textbf{Level type}
\textbf{P}ressure \textbf{l}evel (50-1000 hPa) or \textbf{S}urface  & \textbf{Variable type}: \textbf{P}rognostic, \textbf{D}iagnostic, \textbf{F}orcing & \textbf{Normalization} & \textbf{Scaling} \\
\hline
Geopotential & z & Pl & P & Z-score & 12 \\
\hline
Horizontal wind components & u, v & Pl & P & Z-score & 0.8, 0.5 \\
\hline
Specific humidity & q & Pl & P & Std & 0.6 \\
\hline
Temperature & t & Pl & P & Z-score & 6 \\
\hline
Surface pressure & sp & S & P & Z-score & 10 \\
\hline
Mean sea-level pressure & msl & S & P & Z-score & 1 \\
\hline
Skin temperature & skt & S & P & Z-score & 1 \\
\hline
2 m temperature & 2t & S & P & Z-score & 1 \\
\hline
2 m dewpoint temperature & 2d & S & P & Z-score & 0.5 \\
\hline
10 m horizontal wind components & 10u, 10v & S & P & Z-score & 0.5, 0.5 \\
\hline
Total column water & tcw & S & P & Std & 1 \\
\hline
Volumetric soil water level 1 and 2* & swvl1, swvl2 & S & P & None & 1, 2 \\
\hline
Soil temperature level 1 and 2* & stl1, stl2 & S & P & None & 1, 10 \\
\hline
Total precipitation & tp & S & D & Std & 0.025 \\
\hline
Convective precipitation & cp & S & D & Std (tp) & 0.0025 \\
\hline
Snowfall* & sf & S & D & Std (tp) & 0.025 \\
\hline
Total cloud cover* & tcc & S & D & None & 0.1 \\
\hline
High cloud cover* & hcc & S & D & None & 0.1 \\
\hline
Medium cloud cover* & mcc & S & D & None & 0.1 \\
\hline
Low cloud cover* & lcc & S & D & None & 0.1 \\
\hline
Runoff* & ro & S & D & Std & 0.005 \\
\hline
Surface solar radiation downwards* & ssrd & S & D & Std & 0.05 \\
\hline
Surface thermal radiation downwards* & strd & S & D & Z-score & 0.1 \\
\hline
100 m horizontal wind components* & 100u, 100v & S & D & Z-score & 0.1, 0.1 \\
\hline
Land-sea mask & lsm & S & F & None & \\
\hline
Orography & z & S & F & Max & \\
\hline
Standard deviation of sub-grid orography & sdor & S & F & Max & \\
\hline
Slope of sub-scale orography & slor & S & F & Max & \\
\hline
Insolation & insolation & S & F & None & \\
\hline
Latitude/longitude (cos/sin) & lat/lon & S & F & None & \\
\hline
Time of day/day of year & local time, julian day & S & F & None & \\
\hline
\end{tabular}
\caption{Variables used in the training of AIFS, with their short names, level type, variable type, normalization method, and scaling factors. 
Variables marked with * were newly introduced compared to AIFS v0.2.1.}
\label{tab:aifs_io_variables}
\end{table}

\section{Enforcing Model Constraints}\label{sec:bounding}
Machine-learned forecast models for numerical weather prediction show very good forecast skill, yet they are prone to producing outputs that violate known physical laws or expected statistical consistency. Unlike traditional numerical models, which are governed by equations ensuring mass conservation, positivity, or energy bounds, machine-learned forecast models lack such guarantees by default. As a result, physically implausible outputs, such as negative precipitation, can emerge. We show that incorporating constraints into the model design to enforce physical realism improves forecast skill. In this section, we first identify specific issues in the output of the previous AIFS version related to total precipitation, and then introduce a simple yet effective method to bound the model outputs using activation functions. The proposed method is not restricted to total precipitation but can be equally applied to other variables. 

\subsection{Lack of Physical Realism in Precipitation Forecasts}
\begin{figure}[h]
    \centering
    \includegraphics[width=\linewidth]{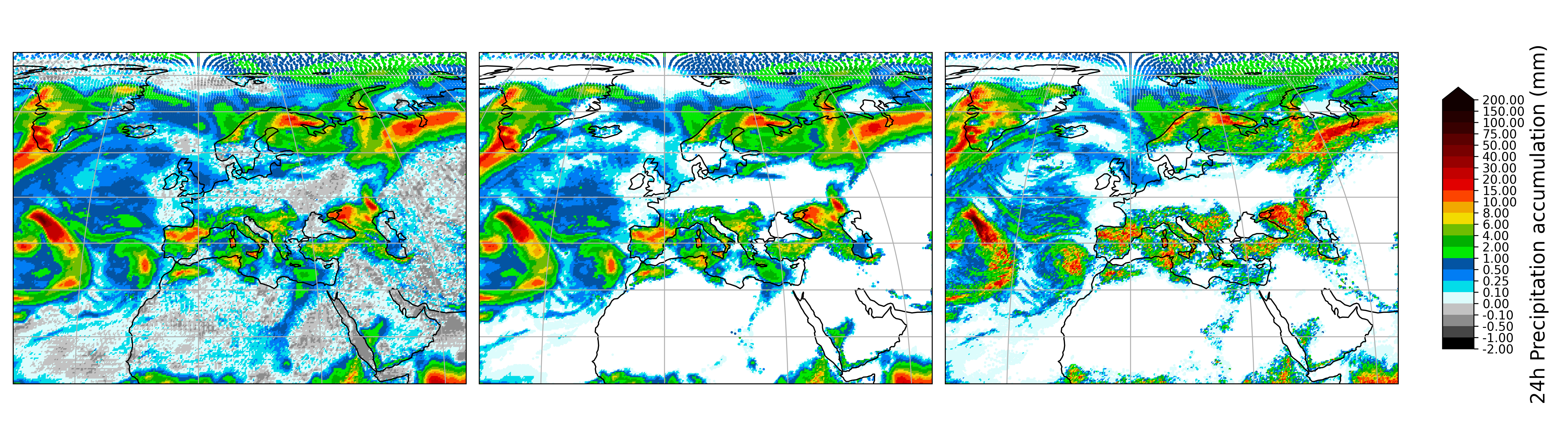}

    \vspace{0 em} 
    \begin{minipage}{\linewidth}
        \centering
        \hspace{-7em} (a) AIFS previous \hspace{5em} (b) AIFS revised \hspace{7em} (c) IFS 
    \end{minipage}
\caption{Comparison of 24-hour total precipitation accumulation from the previous AIFS, the revised AIFS and an estimate derived from the short-range IFS (47r3) 6-hour forecasts, for the forecast issued at 01/06/2023 00:00 UTC and valid at 02/06/2023 00:00 UTC. The previous AIFS shows spurious negative precipitation values and an excess of light rainfall, which are largely corrected in the revised AIFS. The revised version therefore provides a precipitation distribution closer to the IFS reference.}
\label{fig:TP_OLD_VS_NEW}
\end{figure}

The previous AIFS version suffers from significant drawbacks in forecasting precipitation. Most notably, the model's output is not constrained, leading to a frequent occurrence of negative values. This is illustrated in Figure \ref{fig:TP_OLD_VS_NEW}, which compares the 24-hour accumulated total precipitation forecasts from the previous AIFS version, the revised version, and an estimate derived from the short-range IFS (47r3) 6-hour forecasts, for the run initialized on 01/06/2023 at 00:00 UTC and valid at 02/06/2023 00:00 UTC. The previous AIFS shows spurious negative precipitation values and an excess of light rainfall, which are largely corrected in the revised AIFS. While negative values can be clipped to zero at inference time, their presence highlights a lack of physical consistency in the model. This issue is also present in other machine-learned weather forecast models, such as GraphCast \citep{lam2022graphcast}, which is similarly unconstrained.


\begin{figure}[htbp]
    \centering
    \includegraphics[width=0.325\textwidth]{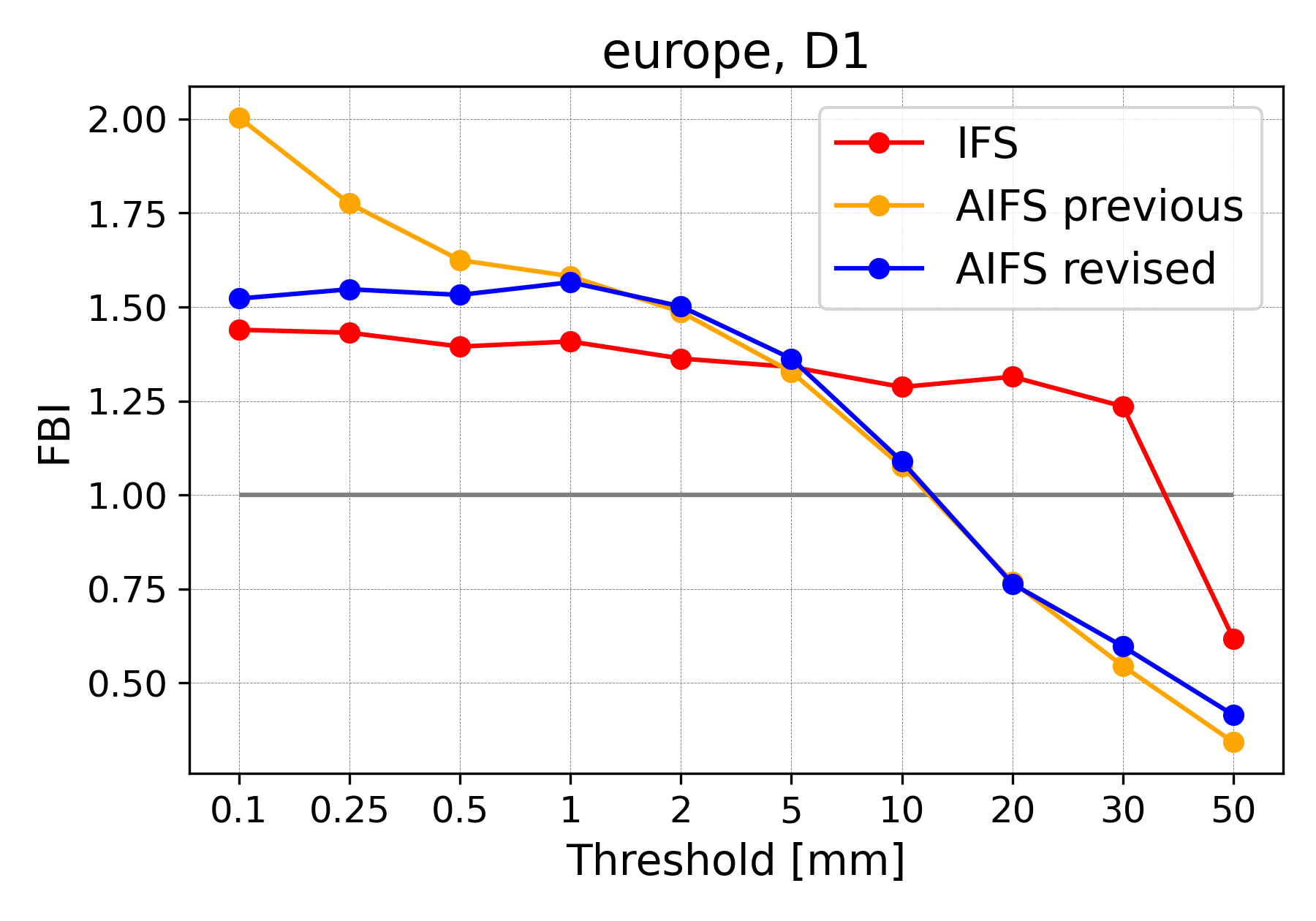}
    \includegraphics[width=0.325\textwidth]{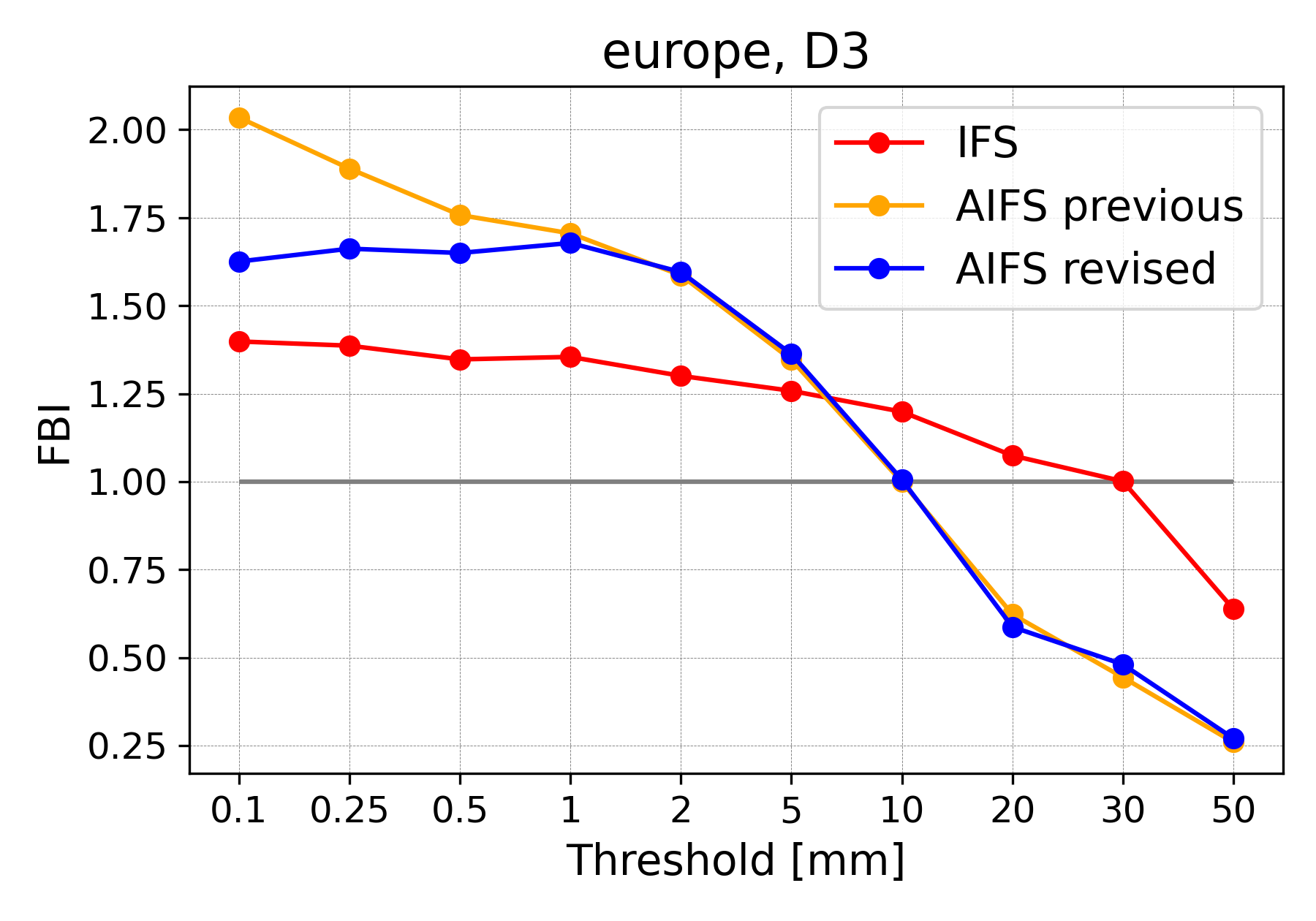}
    \includegraphics[width=0.325\textwidth]{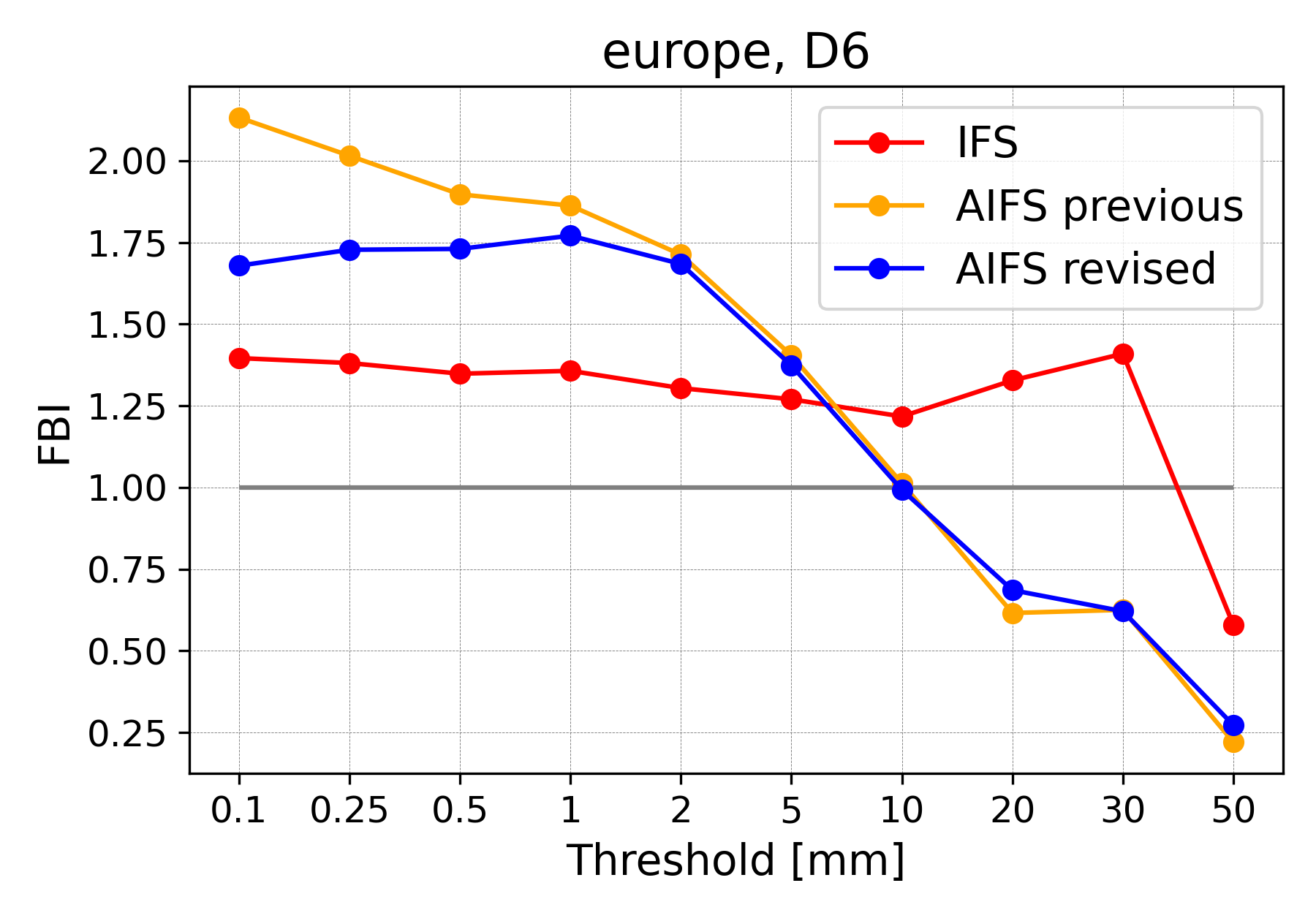}
    
    \begin{subfigure}{0.325\textwidth}
        \centering
        \includegraphics[width=\textwidth]{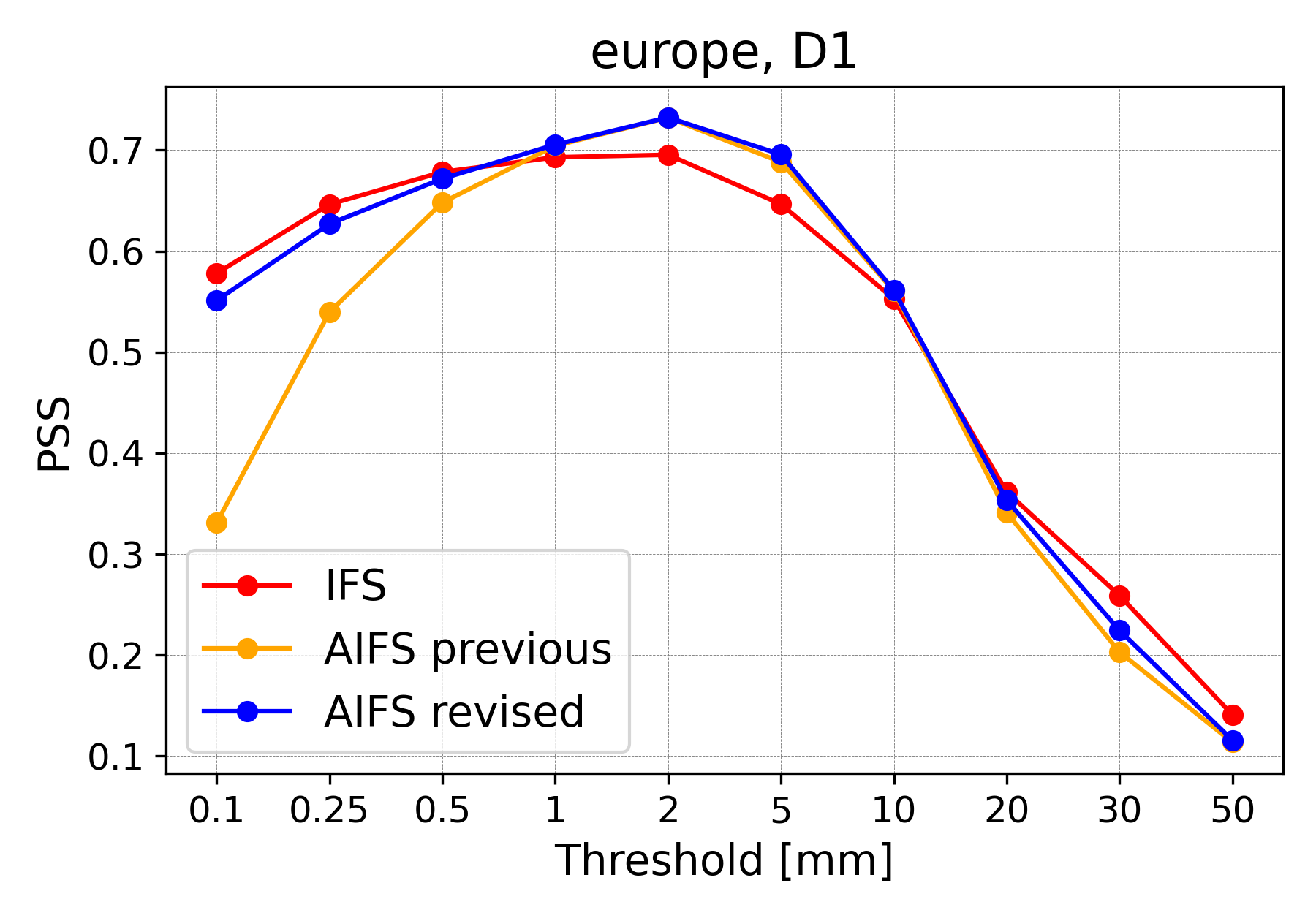}
        \caption{24h}
    \end{subfigure}
    \begin{subfigure}{0.325\textwidth}
        \centering
        \includegraphics[width=\textwidth]{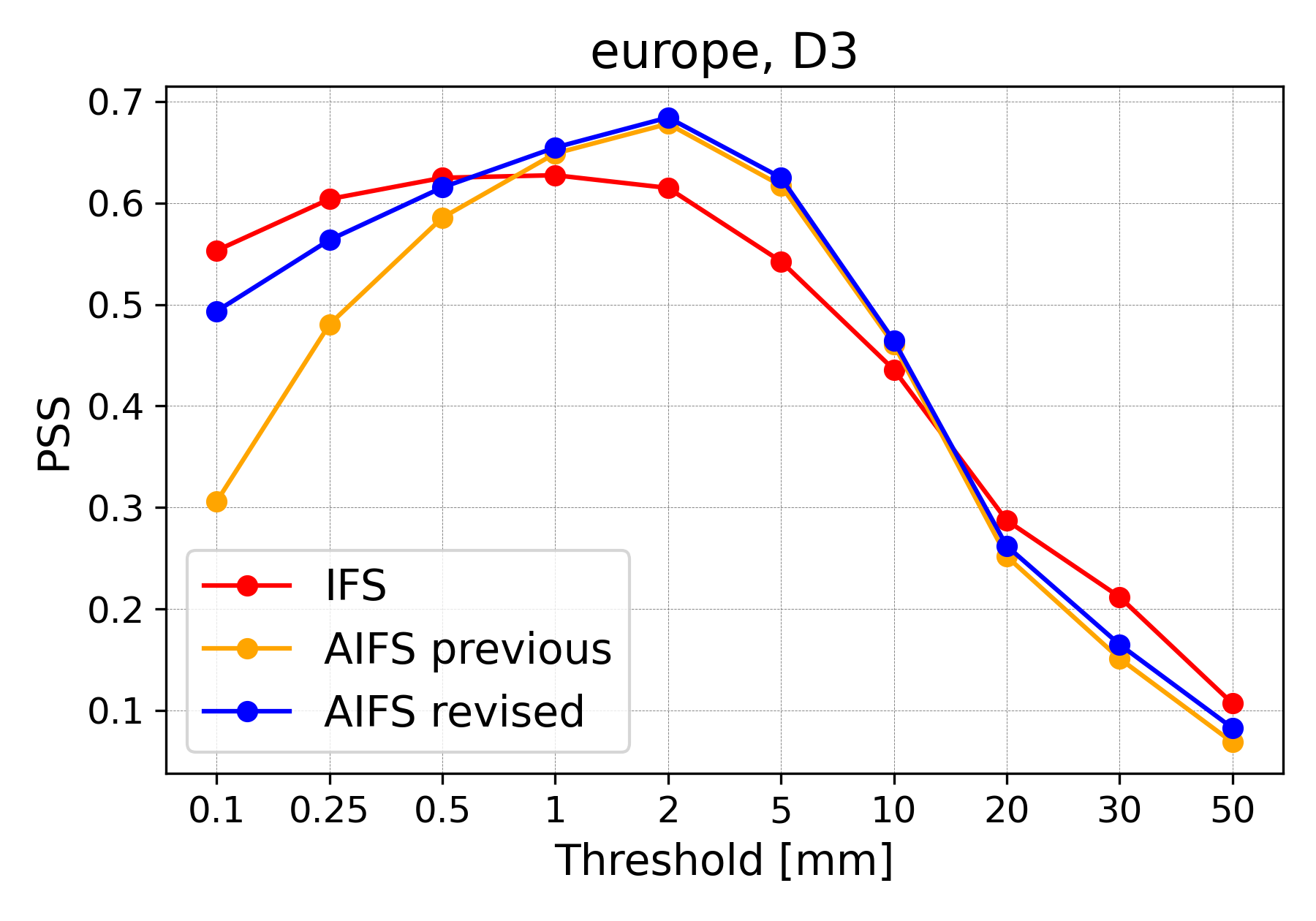}
        \caption{72h}
    \end{subfigure}
    \begin{subfigure}{0.325\textwidth}
        \centering
        \includegraphics[width=\textwidth]{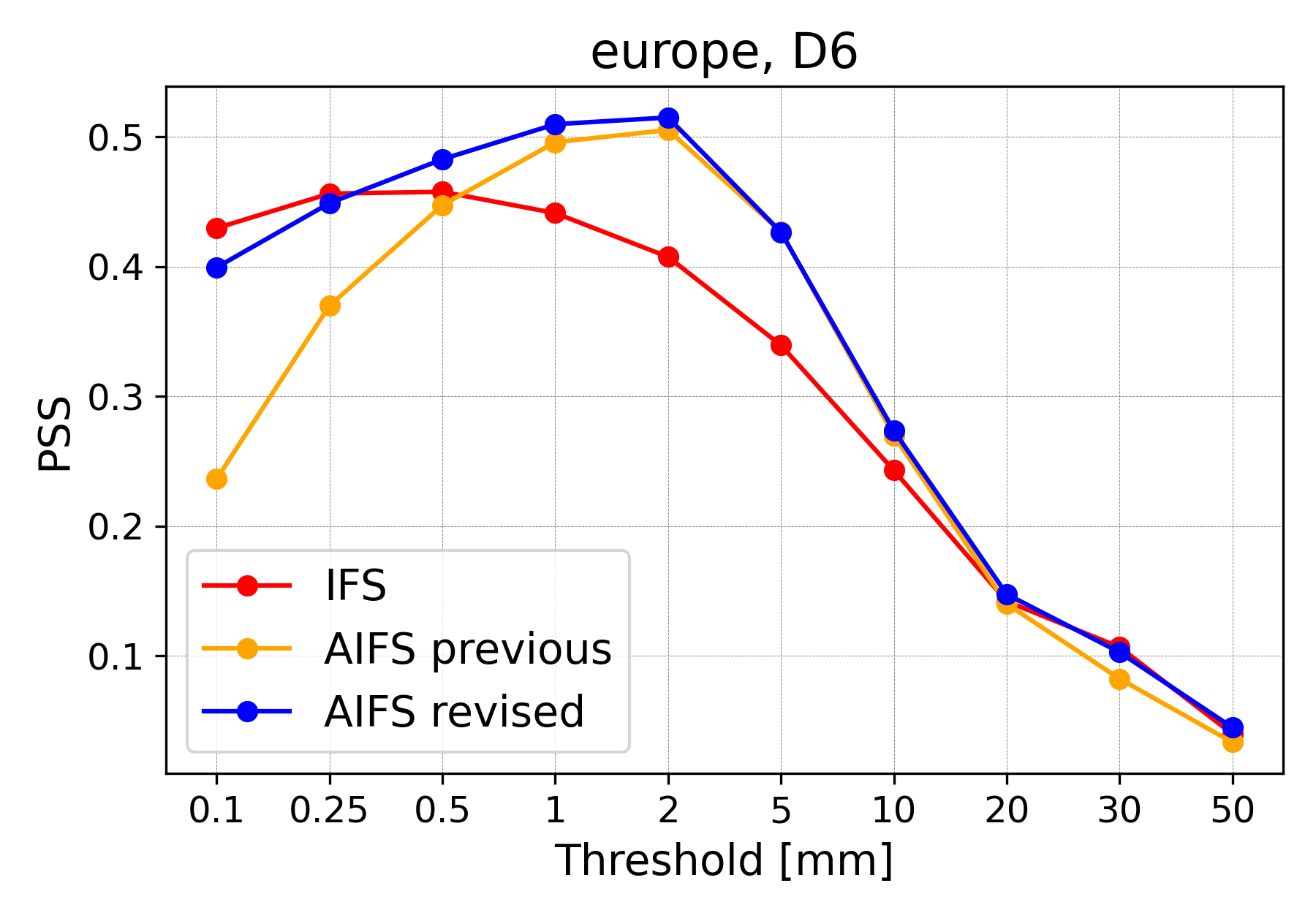}
        \caption{144h}
    \end{subfigure}

    \caption{Comparison of IFS (red), revised AIFS (blue), and previous AIFS (orange) for Europe at forecast steps 24, 72, and 144 hours for 2023. Top row: Frequency Bias Index (FBI); Bottom row: Peirce Skill Score (PSS). The previous version of the AIFS predicts light precipitation in excess.}
    \label{fig:comparison_nhem_ct}
\end{figure}
In addition to the negative values, a second noticeable issue, also visible in Figure \ref{fig:TP_OLD_VS_NEW}, is the excess of light precipitation in the forecast. The model produces excessive light rain leading to a bias in the forecast.

This is further supported by verification metrics computed against in situ observations (SYNOP stations). The Frequency Bias Index (FBI) scores for 2023 over Europe (Figure \ref{fig:comparison_nhem_ct}) confirm that the pre-operational AIFS systematically over-forecasts light precipitation events ($<$ 1 mm). While a similar tendency is present in the IFS, it is considerably more pronounced in the machine-learned forecast model. At the other end of the distribution, the model tends to under-forecast more intense precipitation, as indicated by FBI values well below unity for thresholds exceeding 10 mm. This may be attributed to a well-known characteristic of machine learning-based forecasts: a tendency to produce overly smooth spatial fields, which can suppress extremes. Additionally, the coarser native resolution of AIFS (N320~0.25° grid) compared to IFS (0.1° grid) reduces its spatial representativeness.



\begin{figure}[h]
    \centering
    \includegraphics[width=\linewidth]{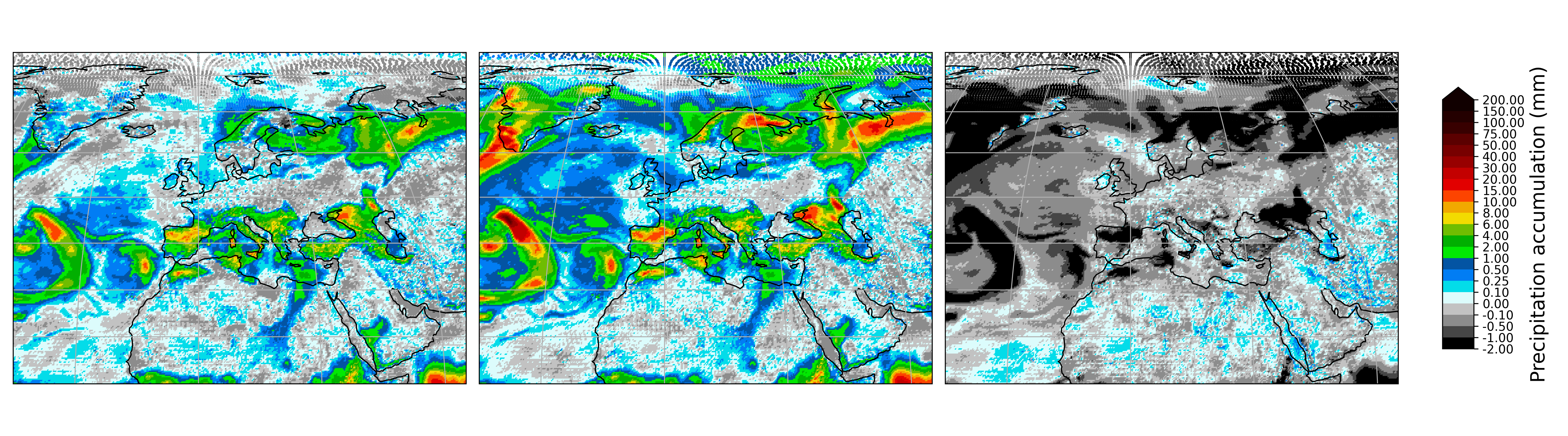}

        \vspace{0 em} 
    \begin{minipage}{\linewidth}
        \centering
        \hspace{-6em} (a) Convective precipitation \hspace{1.5em} (b) Total precipitation \hspace{2.5em} (c) Difference (cp-tp) 
    \end{minipage}
\caption{Comparison of 24-hour total and convective precipitation forecast from the previous AIFS version, together with a map showing the difference between the two of them for the forecast issued at 01/06/2023 00:00 UTC and valid at 02/06/2023 00:00 UTC. Positive values (coloured regions) in the difference plot indicate areas where convective precipitation is greater than the total precipitation.}
\label{fig:TP_CP_old}
\end{figure}

Convective precipitation forecasts also exhibit similar shortcomings. In addition, there is a further lack of physical consistency. Convective precipitation represents the part of the total precipitation that originates from convection, and therefore should always be less than or equal to the total. Figure \ref{fig:TP_CP_old} shows the previous AIFS 24-hour accumulated forecasts of total and convective precipitation for 02/06/2023. The map displaying the difference between the two reveals frequent cases in which convective precipitation exceeds total precipitation, which should not occur. 

\subsection{Bounding the Outputs with Activation Functions}
Precipitation has been used as an example to demonstrate the biases present in the forecasts of some variables. These issues are not only limited to precipitation, but are also observed in all sparsely distributed variables. This behaviour can be avoided by constraining the output of the model.  

There are different strategies one could adopt to enforce physical constraints into the ML model. More specifically, here we tackled unphysical outputs, and we did not consider other constraints such as energy or mass conservation. Introducing loss penalties for outputs that fall outside the known physical bounds can be an effective strategy, and it has the advantage of not requiring any specific model change. Alternatively, the model could be modified in such a way as to prevent output from exceeding variable-specific physical bounds. This is usually referred to as hard-constraining. There are some examples in the literature of hard-constrained machine-learned models for climate and weather, such as \cite{harder2024hardconstraineddeeplearningclimate}. The authors apply a softmax function, a generalization of the logistic function, as a hard-constraint for predicting quantities like atmospheric water content, to enforce the output to be non-negative in climate downscaling. Other examples can be found in \cite{Kent2025-my} or \cite{bonev2025fourcastnet3geometricapproach}. Similarly, we argue that hard constraints on the output can be enforced using an activation function.

Activation functions can be used in a straightforward way to enforce bounds in the output of machine-learned forecast models. Arguably, the most famous activation function and one we used in this work is the Rectified Linear Unit (ReLU), a nonlinear function defined as:
\begin{equation}
\text{ReLU}(x) = \max(0, x)
\end{equation}
ReLU maps all negative values to zero, effectively enforcing a hard lower bound on the output.  
For variables requiring both upper and lower bounds, such as concentrations or fractions, the Hard Hyperbolic Tangent (HardTanh) function is an effective choice. It is a piecewise linear approximation of the hyperbolic tangent, defined as:

\[
\text{HardTanh}(x) = 
\begin{cases}
0 & \text{if } x < 0 \\
x & \text{if } 0 \leq x \leq 1 \\
1 & \text{if } x > 1 . 
\end{cases}
\]

HardTanh can also be used to enforce consistency between related output variables. For instance, consider the case of convective precipitation (Figure \ref{fig:TP_CP_old}), which is predicted independently of total precipitation in the previous AIFS version. There is a clear relation between the two quantities: convective precipitation is a fraction of total precipitation and should never exceed it. A more physically consistent approach is to map the original convective output to the [0,1] range using a HardTanh layer and to multiply this output by the predicted total precipitation:

\begin{equation}
    \text{cp} = \text{HardTanh}(\text{cp}^{'}) \times \text{tp},
\end{equation}

where $\text{cp}^{'}$ is the convective precipitation output before the activation layer. This guarantees consistency. This type of constraint, referred to as FractionBounding, is applied to variables related to total precipitation and total cloud cover.

Clipping the precipitation output in inference is a possibility and a common practice. This was the case in the pre-operational AIFS model and also reported in other studies, such as \cite{balogh2024onlinetestneuralnetwork}. However, we show that the introduction of bounding in the output during training has benefits beyond simply avoiding slightly negative or unphysical values: it can facilitate the learning of forecasting for sparse and intermittent variables. Bounding effectively decomposes the prediction space into two distinct regions. In the case of total precipitation, the negative space becomes a proxy for forecasting the non-event, while the positive space corresponds to the occurrence of precipitation. This decomposition may, in principle, help the model more easily perform a classification between event and non-event outcomes, a distinction the previous AIFS version struggles with.

Table \ref{tab:bounding-strategies} summarises the bounding strategy used in the new version of the AIFS. Since bounding is performed on the normalized space, the choice of the normalization strategy is essential. In particular, variables bounded using a ReLU function were normalized using the standard deviation only, as indicated in Table \ref{tab:aifs_io_variables}, to avoid offsetting the zero value. Since snowfall and convective precipitation are predicted as fractions of total precipitation, it is necessary to ensure consistent magnitudes in the normalized space. Therefore, cp and sf were scaled using the standard deviation of total precipitation rather than their own. Total cloud cover and soil moisture variables (swvl1 $\&$ swvl2) were not normalized, since their range falls within the constraints imposed by the HardTanh bounding ([0,1]).

\begin{table}[ht]
\centering
\begin{tabular}{|l|l|p{7cm}|}
\hline
\textbf{Bounding Type} & \textbf{Range} & \textbf{Variables} \\
\hline
ReluBounding & [0, $\infty$) & 
tp, ro, tcw, ssrd, q(50-1000 hPa) \\
\hline
HardtanhBounding & [0, 1] & 
tcc, swvl1, swvl2 \\
\hline
FractionBounding (w.r.t. tp) & [0, 1] & 
cp, sf \\
\hline
FractionBounding (w.r.t. tcc) & [0, 1] & 
lcc, mcc, hcc \\
\hline
\end{tabular}
\caption{Summary of bounding strategies used in the new version of AIFS. }
\label{tab:bounding-strategies}
\end{table}

\section{Evaluation}

\begin{figure}[htbp]
    \centering
    \begin{subfigure}[t]{0.49\linewidth}
        \includegraphics[width=\linewidth, page=1]{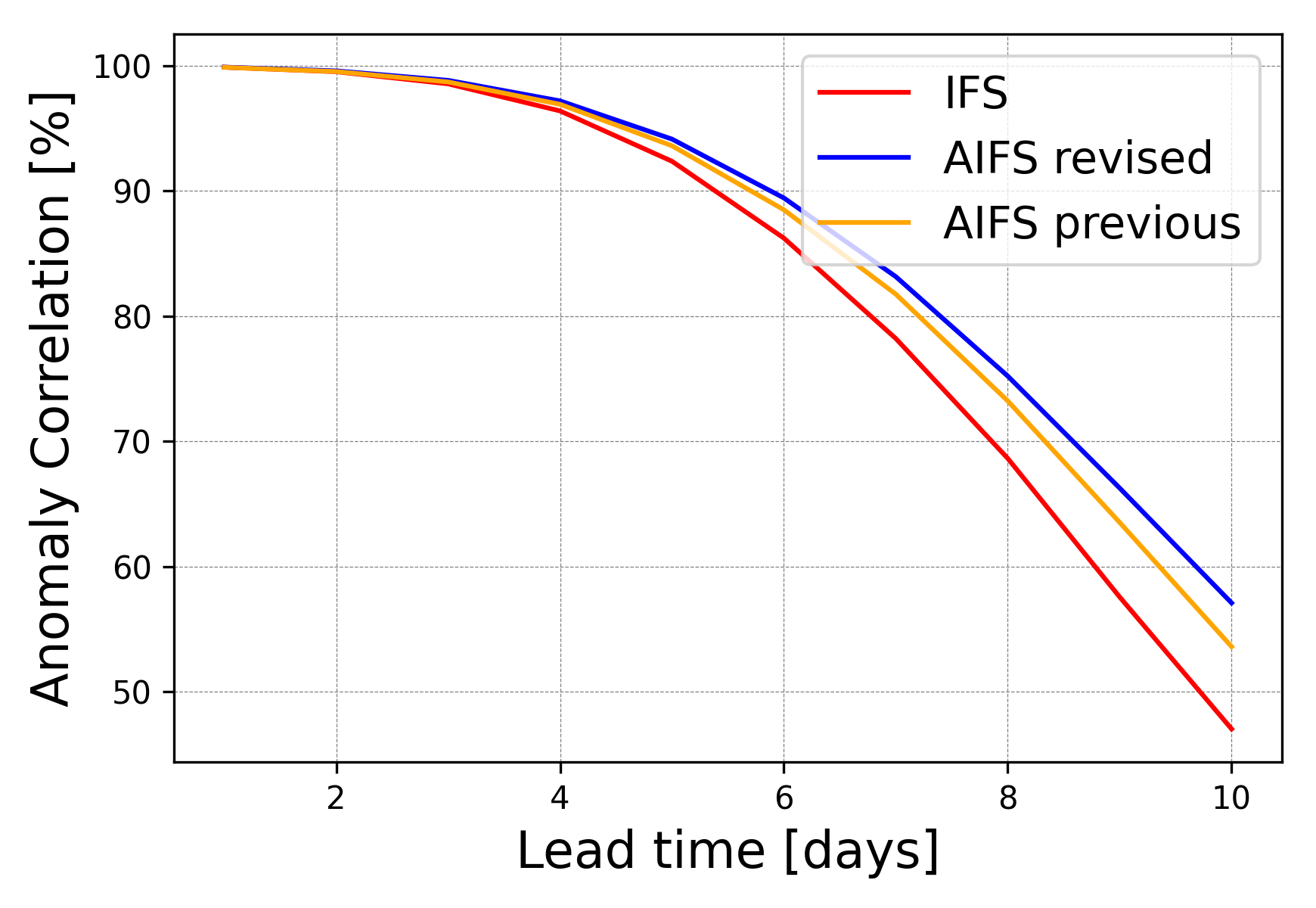}
        \caption{Geopotential at 500hPa}
    \end{subfigure}
    \hfill
    \begin{subfigure}[t]{0.49\linewidth}
        \includegraphics[width=\linewidth, page=2]{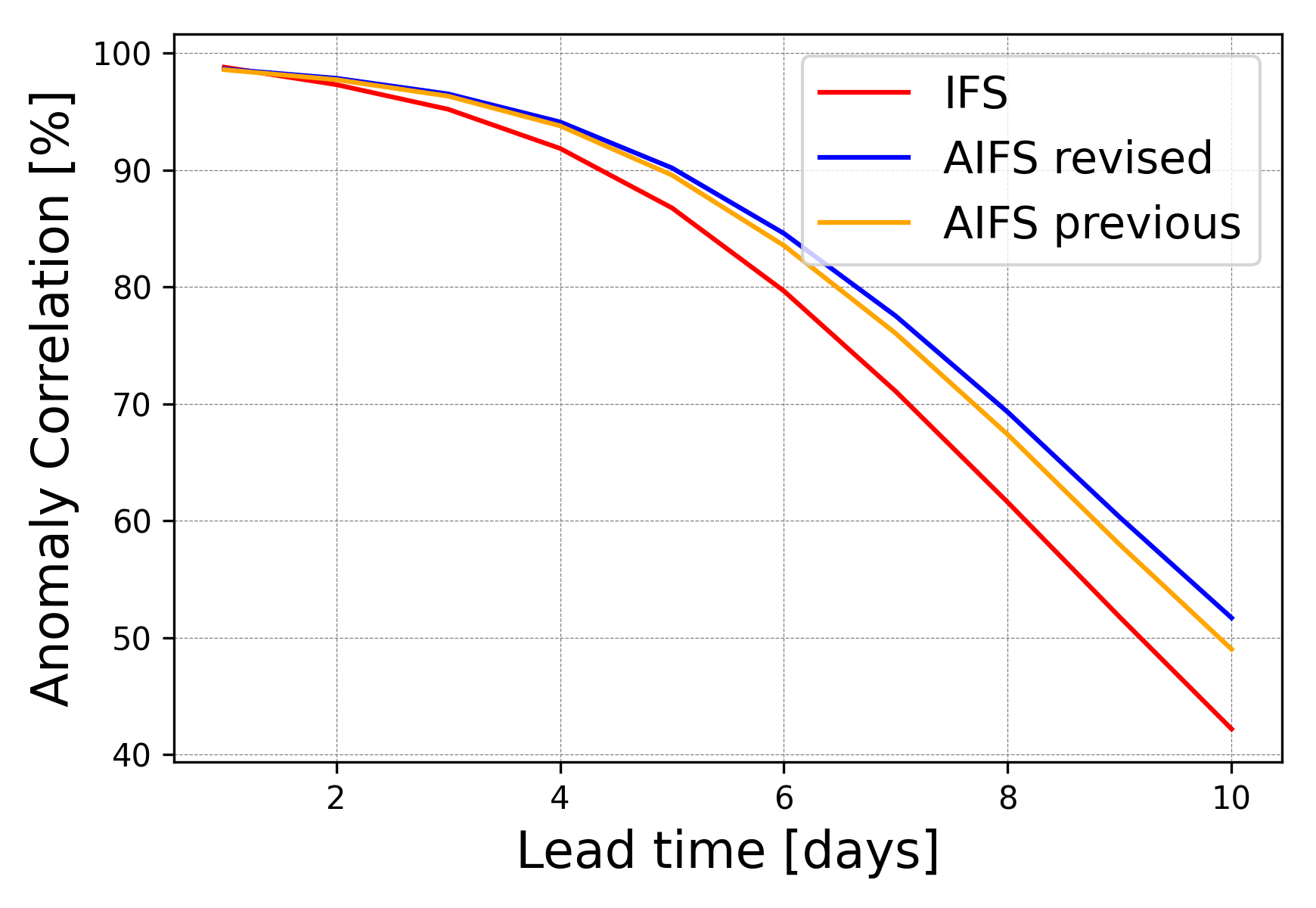}
        \caption{Temperature at 850hPa}
    \end{subfigure}
    \caption{Anomaly correlation skill scores for geopotential and temperature at 500hPa and 850hpa, respectively. Skill scores computed for the Northern Hemisphere for the whole of 2023 against IFS analysis. In the medium range, AIFS revised outperforms the IFS by 12 to 24 hours in skill. Forecast skill is also clearly improved compared to the
previous AIFS version.}
    \label{fig:z500 scores}
\end{figure}

\begin{figure}[htbp]
    \centering
    \begin{subfigure}[t]{0.49\linewidth}
        \includegraphics[width=\linewidth, page=4]{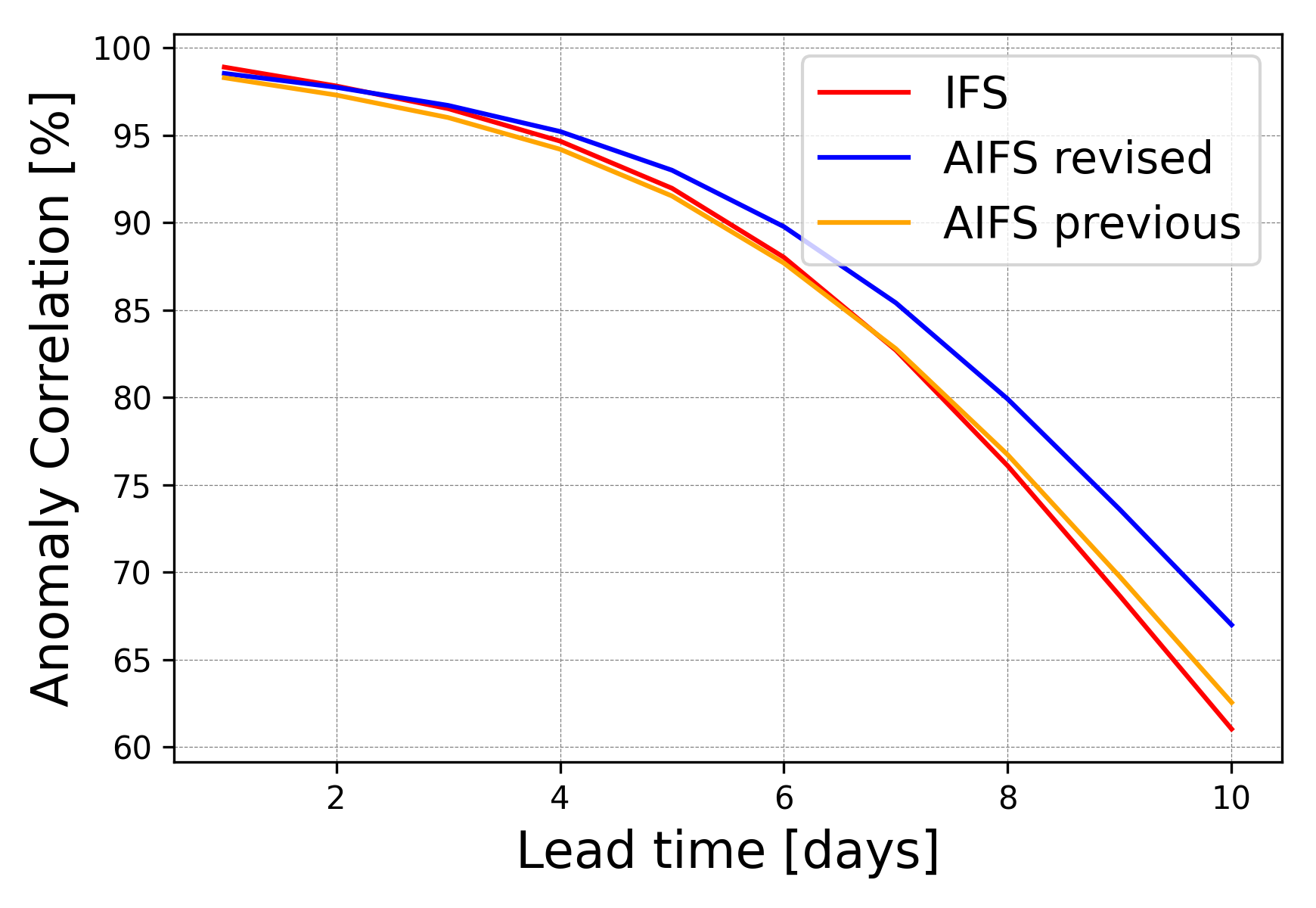}
        \caption{Temperature at 100hPa}
    \end{subfigure}
    \hfill
    \begin{subfigure}[t]{0.49\linewidth}
        \includegraphics[width=\linewidth, page=7]{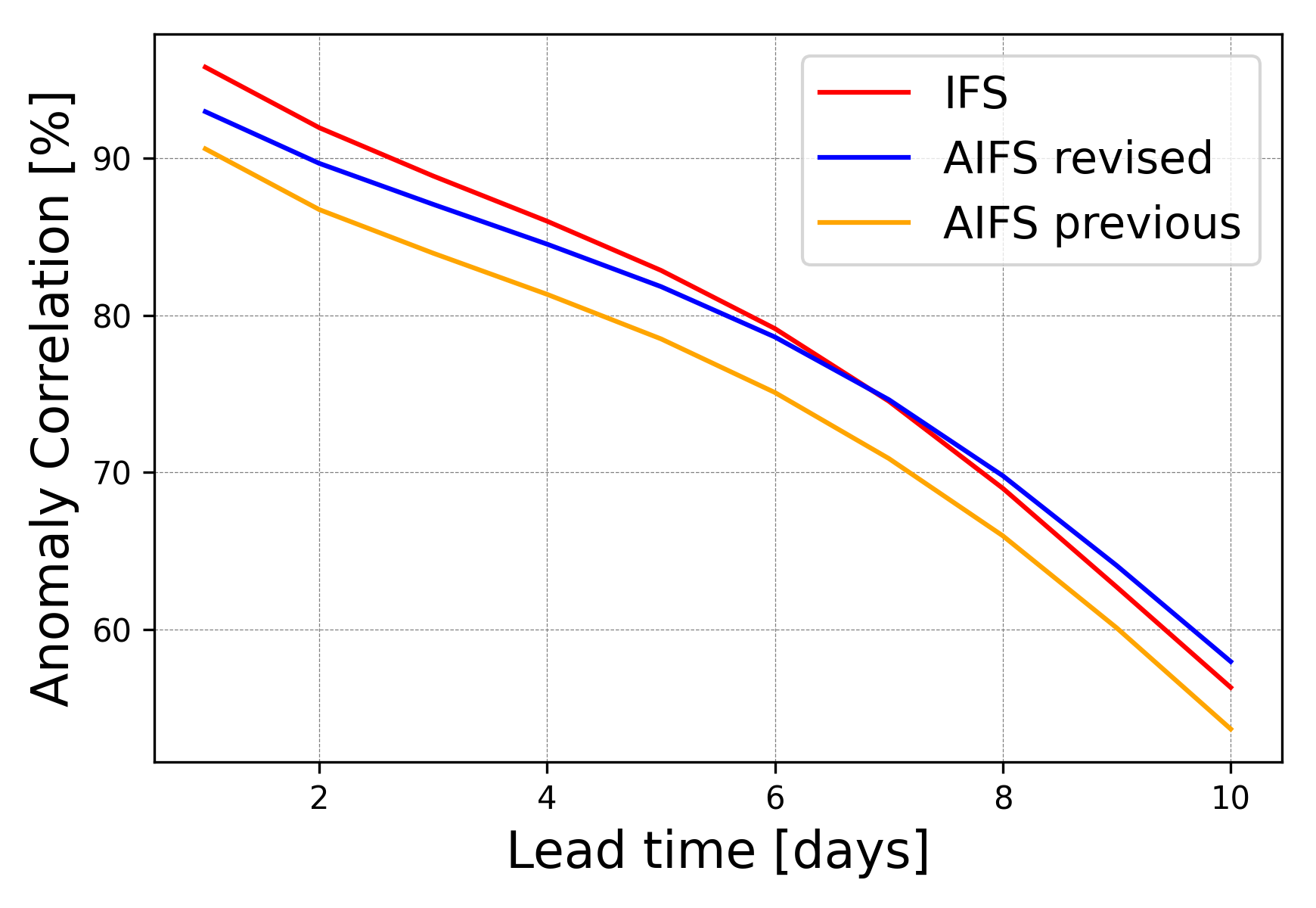}
        \caption{Wind Speed at 50hPa}
    \end{subfigure}
    \caption{Anomaly correlation skill scores for temperature at 100hPa and wind speed at 50hPa. Skill scores computed for the Northern Hemisphere for the whole of 2023 against IFS analysis. Significant improvements in the revised AIFS forecasts at 100 and 50 hPa when compared against the previous AIFS version.}
    \label{fig:stratosphere_scores}
\end{figure}

The revised AIFS version delivers highly skilled forecasts, as shown by anomaly correlation scores for 2023 in the Northern Hemisphere (Figure \ref{fig:z500 scores}). In the medium range, AIFS outperforms the IFS by 12 to 24 hours in skill. Forecast skill is also clearly improved compared to the previous AIFS version. 
This can be attributed to more training data and improvements in rollout fine-tuning. Here, we verify against the operational IFS analysis, which is also used to initialise the forecasts. Additionally, imposing a minimum on the loss weights in the stratosphere leads to significant improvements in the data-driven forecasts at 100 and 50 hPa (Figure \ref{fig:stratosphere_scores}). For temperature at 100hPa, the new version of the AIFS outperforms the IFS, while for 50hPa wind speed, the gap in skill between the previous version of AIFS and the IFS in the stratosphere is significantly reduced.

\begin{figure}[h]
    \centering
    \includegraphics[width=0.68\linewidth]{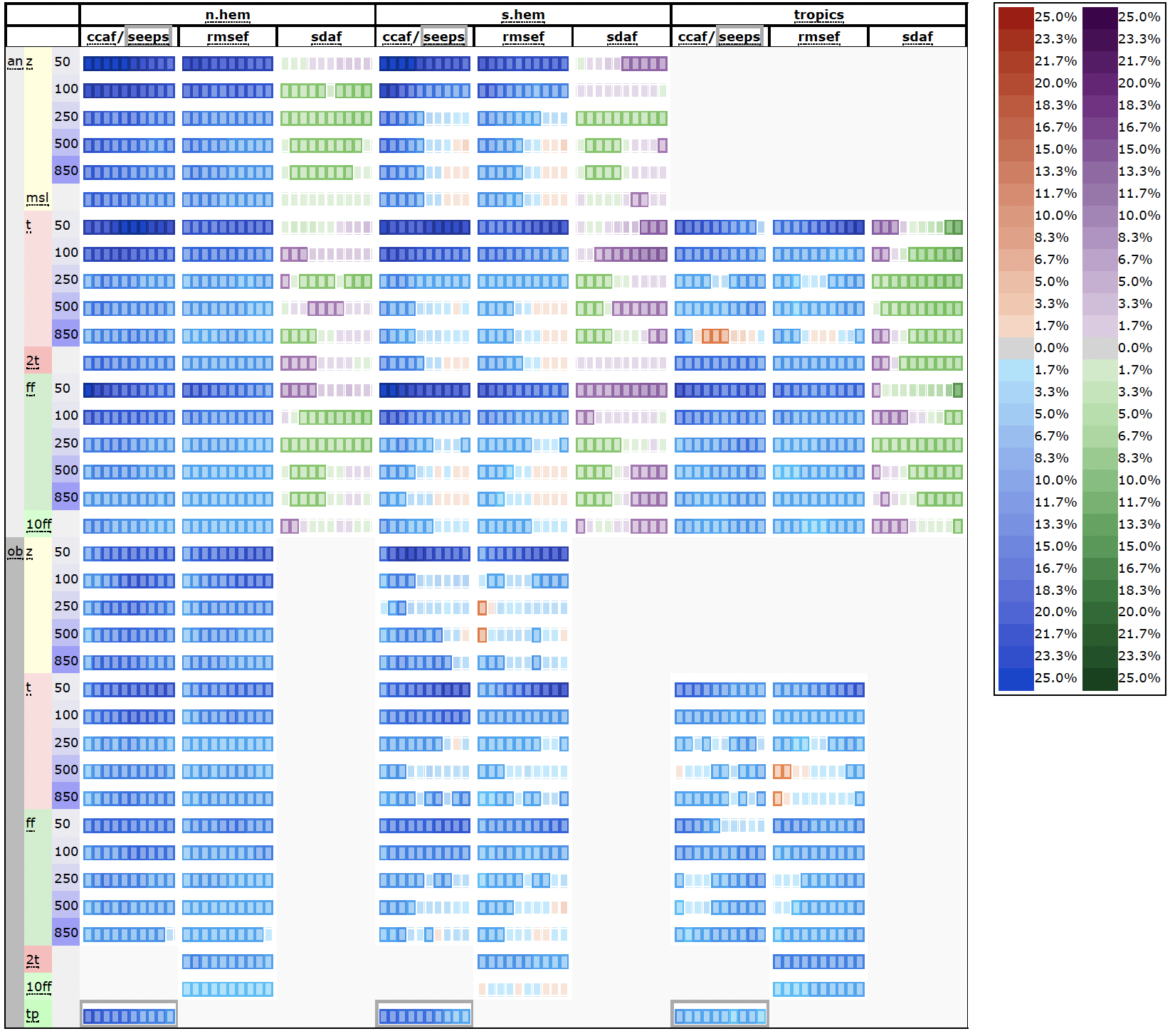}
\caption{Scorecard comparing forecast scores of AIFS revised versus the previous AIFS version for the whole year of 2023. Forecasts are initialised on 00 and 12 UTC. Relative score changes are shown as function of lead time (day 1 to 10) for northern extra-tropics (n.hem), southern extra-tropics (s.hem) and tropics. Blue colours mark score improvements and red colours score degradations. Purple colours indicate an increased in standard deviation of forecast anomaly, while green colours indicate a reduction. Framed rectangles indicate 95$\%$ significance level. Numbers behind variable abbreviations indicate variables on pressure levels (e.g., 500~hPa), and suffix indicates verification against IFS NWP analyses (an) or radiosonde and SYNOP observations (ob). Scores shown are anomaly correlation (ccaf), SEEPS (seeps, for 24h precipitation accumulation),  RMSE (rmsef) and standard deviation of forecast anomaly (sdaf).}
\label{fig:scorecard}
\end{figure}

\begin{figure}[htbp]
    \centering
    \begin{subfigure}[t]{0.49\linewidth}
        \includegraphics[width=\linewidth]{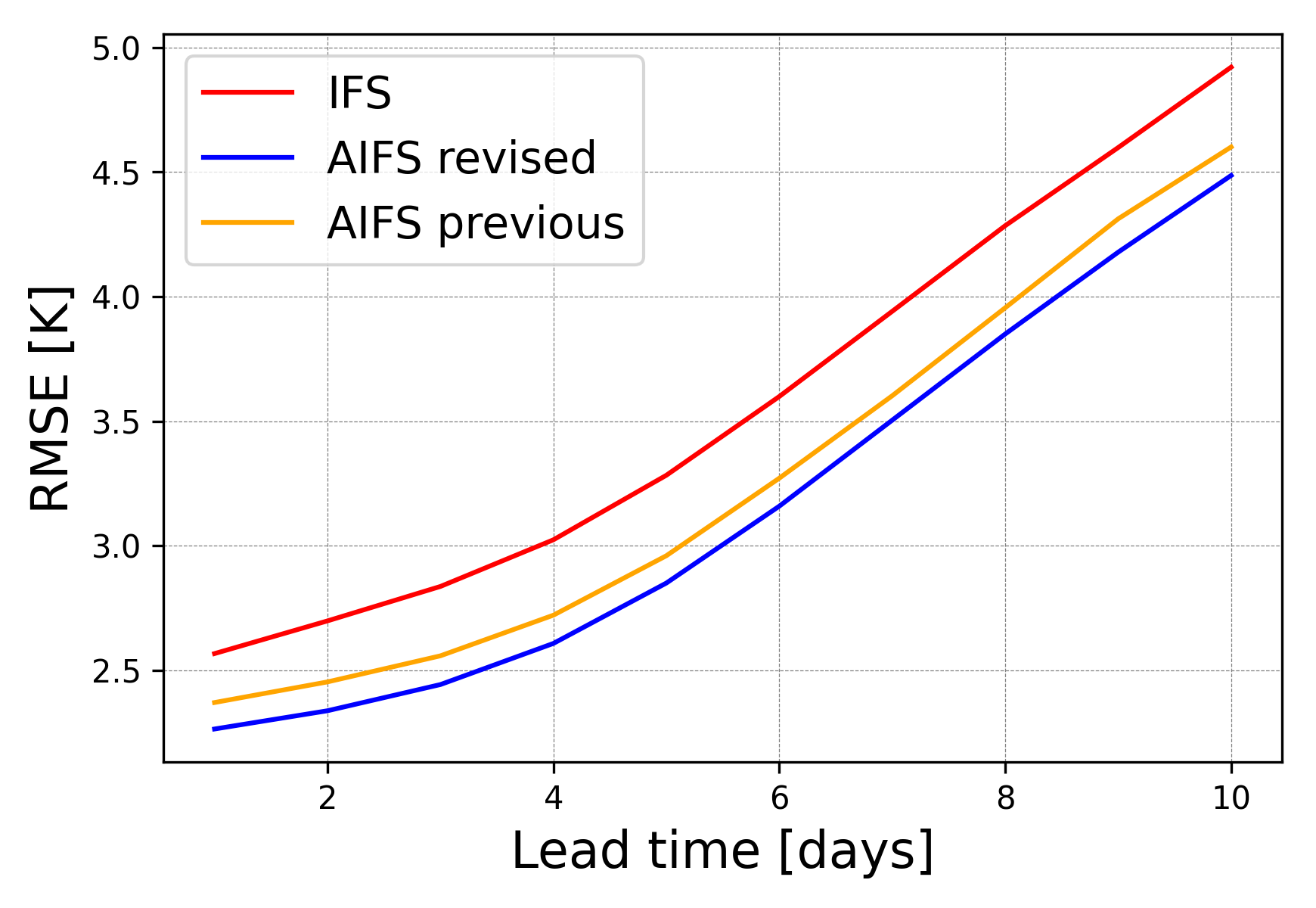}
        \caption{2-metre temperature}
    \end{subfigure}
    \hfill
    \begin{subfigure}[t]{0.49\linewidth}
        \includegraphics[width=\linewidth, page=1]{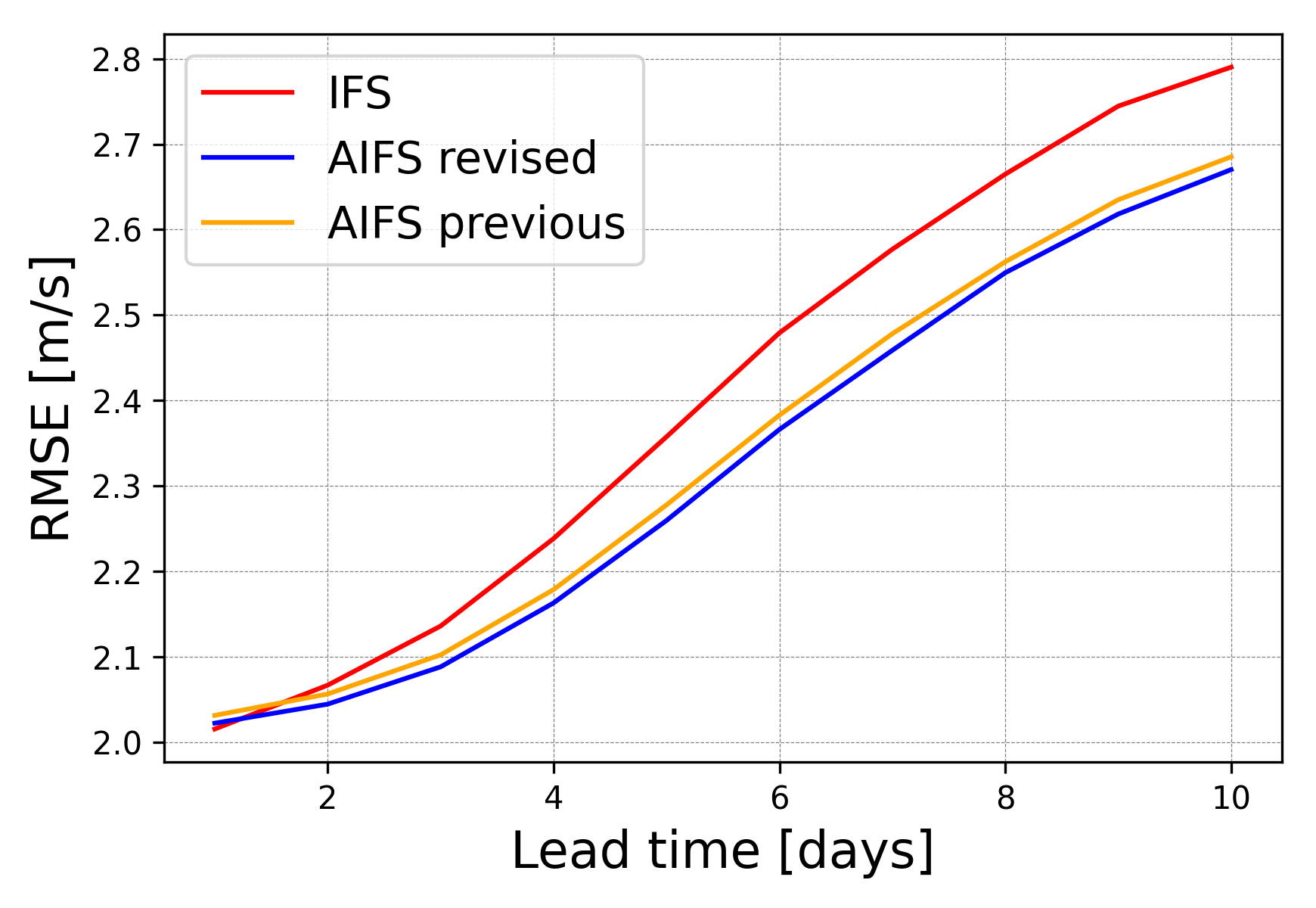}
        \caption{10-metre wind speed}
    \end{subfigure}
    \caption{RMSE scores for 2-metre temperature and 10-metre wind speed computed against SYNOP observations over the Northern Hemisphere. The revised AIFS version shows improvement when compared to the previous verision of the AIFS.}
    \label{fig:t2m_10ff_rmse}
\end{figure}

Forecast skill for key surface variables, such as 2-metre temperature and 10-metre wind speed, verified against SYNOP observations, is similarly improved (Figure \ref{fig:t2m_10ff_rmse}). Overall, the new AIFS version exhibits improvements of around 4–6~$\%$ across all variables, lead times, and pressure levels relative to the previous AIFS version, as shown in the scorecard presented in Figure \ref{fig:scorecard}. The performance of the model for tropical cyclone prediction is similar to that of the previous version (see \citet{lang2024aifs}), with some small improvements to track position. As a design choice, rollout fine-tuning was configured to ensure that the field smoothness characteristics remain consistent with those of the previous AIFS version. This was confirmed by spectral analysis (not shown here).

\begin{figure}[htbp]
    \centering

    \begin{subfigure}[b]{0.325\textwidth}
        \centering
        \includegraphics[width=\linewidth,page=3]{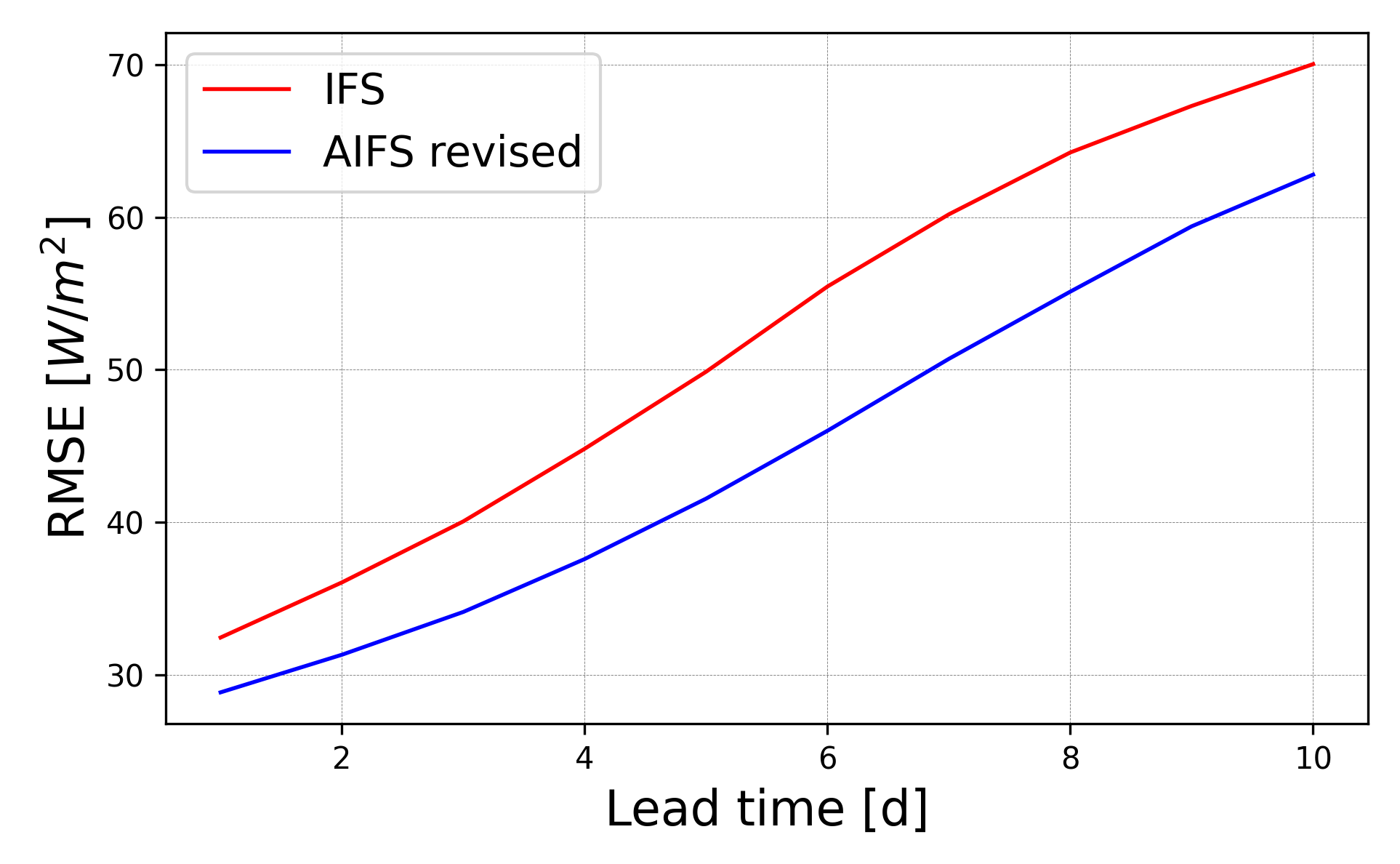}
        \caption{ssrd}
    \end{subfigure}
    \begin{subfigure}[b]{0.32\textwidth}
        \centering
        \includegraphics[width=\linewidth, page=4]{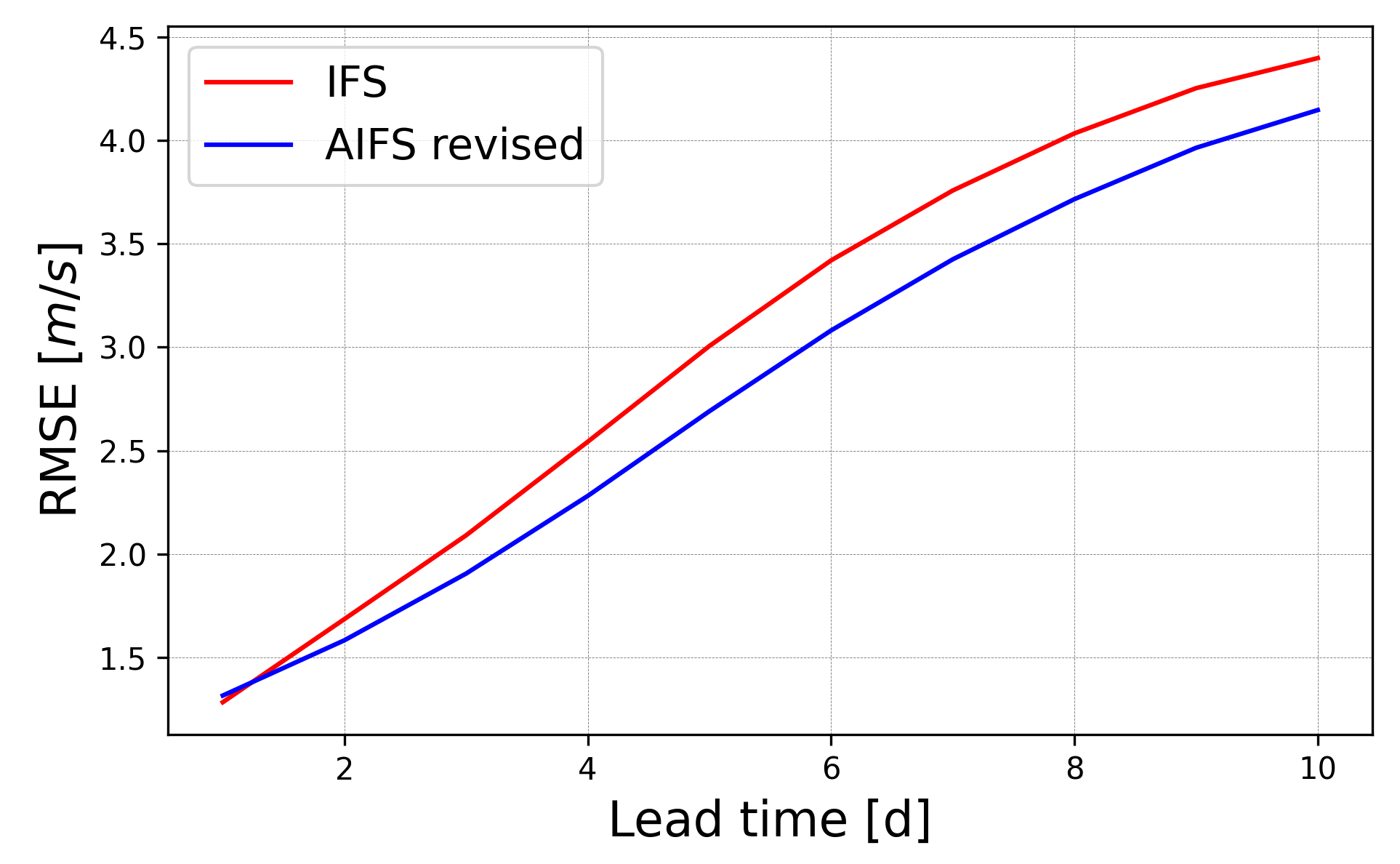}
        \caption{100-metre winds}
    \end{subfigure}
    \begin{subfigure}[b]{0.325\textwidth}
        \centering
        \includegraphics[width=\linewidth, page = 5]{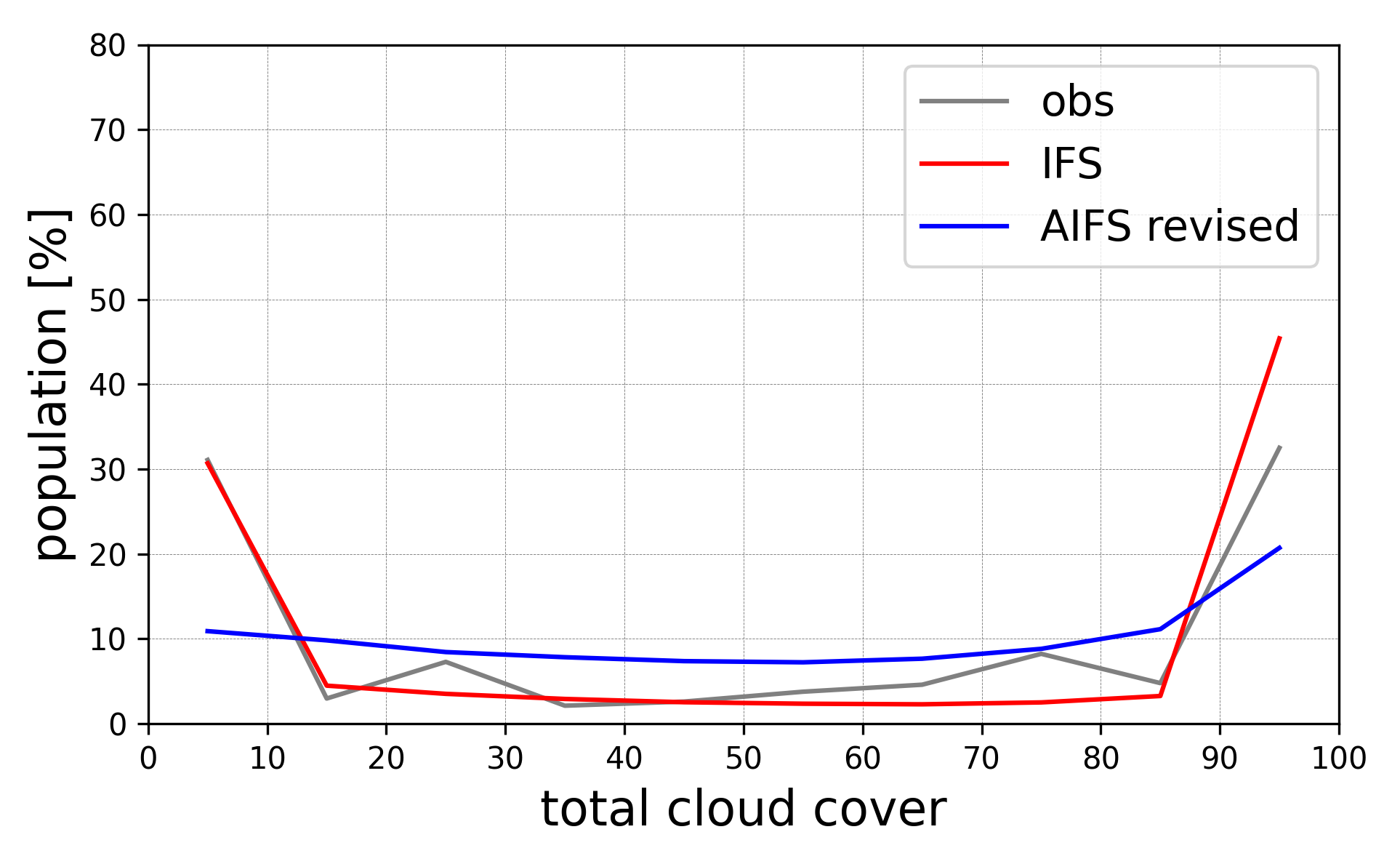}
        \caption{total cloud cover, day 5}
    \end{subfigure}

    \caption{Forecast RMSE computed against operational IFS analysis and distribution comparison for new variables. (a) Surface solar radiation downwards RMSE for March–May (MAM) 2023, (b) 100-metre wind speed RMSE for the full year 2023, (c) Total cloud cover distribution for June–August (JJA) 2023.  Blue lines show the AIFS revised and red lines show IFS; observations are shown in grey in panel~(c). AIFS shows significant gains in forecast skill in the medium range for surface short-wave downwards radiation and 100-metre winds when compared against the IFS. The mismatch in population distribution for total cloud cover forecast highlights the inherent limitations of MSE-trained AI models.}
    \label{fig:eval_tcc_solrad_100mwinds}
\end{figure}

Figure \ref{fig:eval_tcc_solrad_100mwinds} presents verification metrics for several variables introduced in the new version. In line with those already present in earlier versions, AIFS shows a gain in forecast skill of around one day in the medium range for surface short-wave downwards radiation verified against geostationary satellite observation via CMSAF \citep{https://doi.org/10.5676/eum_saf_cm/sarah/v003} and 100-metre wind speed verified against ECMWF operational analysis, relative to the IFS. The population distribution for total cloud cover verified against SYNOP observations, however, highlights the inherent limitations of MSE-trained AI models. While the observed distribution follows a U-shape, with high frequency at the tails of the distribution (clear skies and overcast conditions), AIFS produces a much flatter distribution, under-predicting these extremes and over-estimating intermediate values. This behaviour is closely linked to the smoothing effect introduced by the MSE loss function, which tends to penalize large deviations and thereby suppress extremes (see Section \ref{sec:discussion}).

\begin{figure}[htbp]
    \centering

    \begin{subfigure}[b]{0.495\textwidth}
        \centering
        \includegraphics[width=\linewidth]{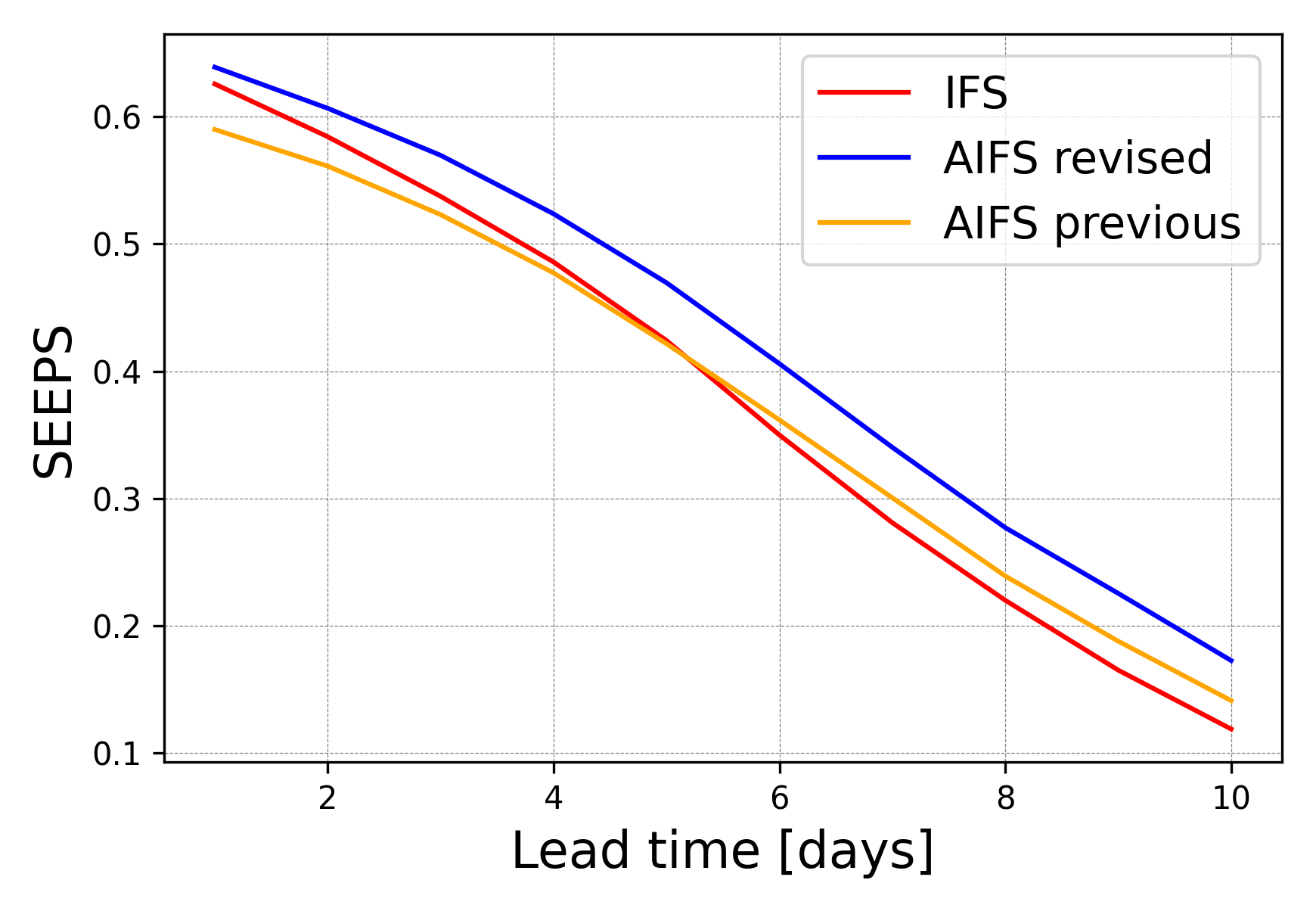}
        \caption{Northern Hemisphere }
    \end{subfigure}
    \begin{subfigure}[b]{0.495\textwidth}
        \centering
        \includegraphics[width=\linewidth]{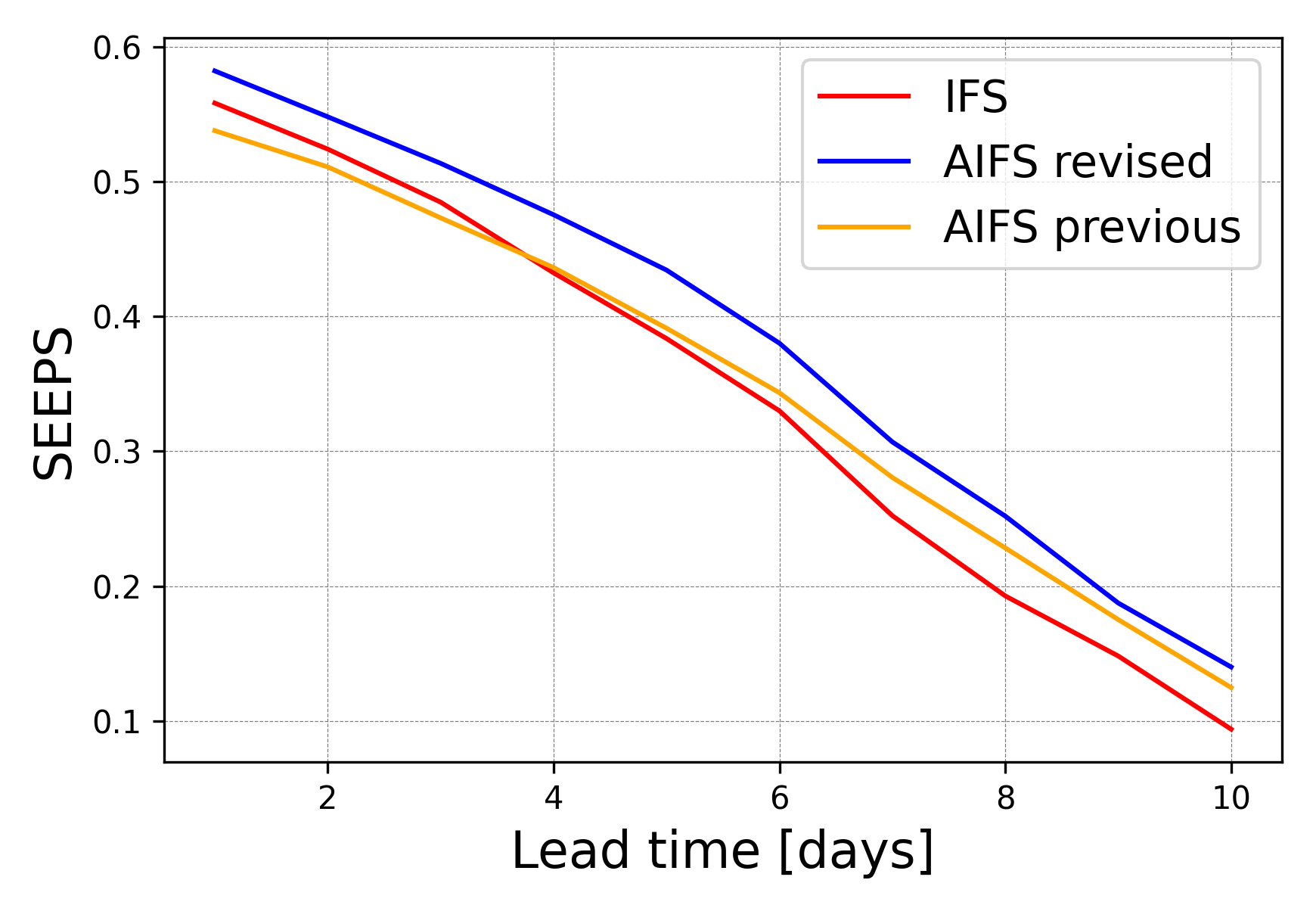}
        \caption{Southern Hemisphere}
    \end{subfigure}

    \caption{SEEPS skill scores for 2023 based on 24-hour accumulated precipitation from SYNOP observations, comparing the revised AIFS (blue), the previous AIFS version (orange), and the IFS (red) across different regions. Results show a consistent and statistically significant improvement across all lead times and in the three main global regions for the revised AIFS version when compared to the previous AIFS version and the IFS.}

    \label{fig:seeps_scores}
\end{figure}

The forecasting skill of the model with respect to 24-hour accumulated total precipitation is significantly improved. The new AIFS version is compared against both the previous AIFS version and the operational IFS (cycles 47r3 and 48r1) in Figure \ref{fig:seeps_scores}. The Stable Equitable Error in Probability Space (SEEPS) skill score (\cite{rodwellseeps}) is used as the primary verification metric, with 24-hour accumulated precipitation SYNOP observations serving as the reference. Results show a consistent and statistically significant improvement across all lead times and in the three main global regions: the Northern Hemisphere, the Southern Hemisphere, and the tropics. The revised AIFS demonstrates approximately a one-day gain in forecast skill relative to both IFS and the previous AIFS version. The forecast fields also exhibit noticeable improvements, as illustrated in Figure \ref{fig:TP_OLD_VS_NEW}. The new version of the AIFS produces no negative values in the output and substantially reduces light precipitation, aligning more closely with the 24-hour total precipitation accumulation fields derived from the IFS operational short-range forecasts.

Figure \ref{fig:comparison_nhem_ct} reveals where the improvement originates. The Frequency Bias Index (FBI) and Peirce Skill Score (PSS) are shown for the Northern Hemisphere for different thresholds. The previous AIFS version exhibits a strong tendency to over-predict light precipitation events ($<1$ mm) across all lead times, as shown by the FBI. This bias is substantially corrected due to the bounding (see Section \ref{sec:b_tp}) in the revised AIFS.

While the AI model still slightly over-predicts light precipitation compared to the IFS, it demonstrates competitive skill for light precipitation. The AIFS excels at medium-intensity events (1–10 mm), with PSS scores significantly higher than those of the IFS. At higher thresholds ($>$ 10mm), corresponding to moderate to heavy precipitation, the AIFS diverges from the IFS, with a marked under-prediction (FBI $<1$). This is likely caused by smoothing introduced by the loss function, in combination with the model’s coarser spatial resolution.

This under-prediction plays an important role in the metrics concerning more extreme events, since both the previous and the revised AIFS models underperform IFS for thresholds exceeding 10mm in terms of PSS, but remains competitive. This suggests that although the AI models predict fewer high-intensity events, their predictions are more accurate when they do occur. Finally, the revised AIFS shows a marginal improvement in terms of PSS compared against the previous AIFS version, possibly due to improvements in the learning-rate scheduling used for fine-tuning and additional training data.

\subsection{Evaluating the effects of bounding on total precipitation} \label{sec:b_tp}
Overall, the revised AIFS version demonstrates significant improvements in forecasting skill for total precipitation over its predecessor. The bounding of total precipitation transforms the prediction space such that negative values correspond to ``no-rain'' and positive values to ``rain''. This separation enables the model to more effectively distinguish between the two scenarios. It removes the pressure to forecast exactly zero and facilitates the classification task inherent to precipitation forecasting.

\begin{figure}[htbp]
    \centering

    \begin{subfigure}[b]{0.495\textwidth}
        \centering
        \includegraphics[width=\linewidth,]{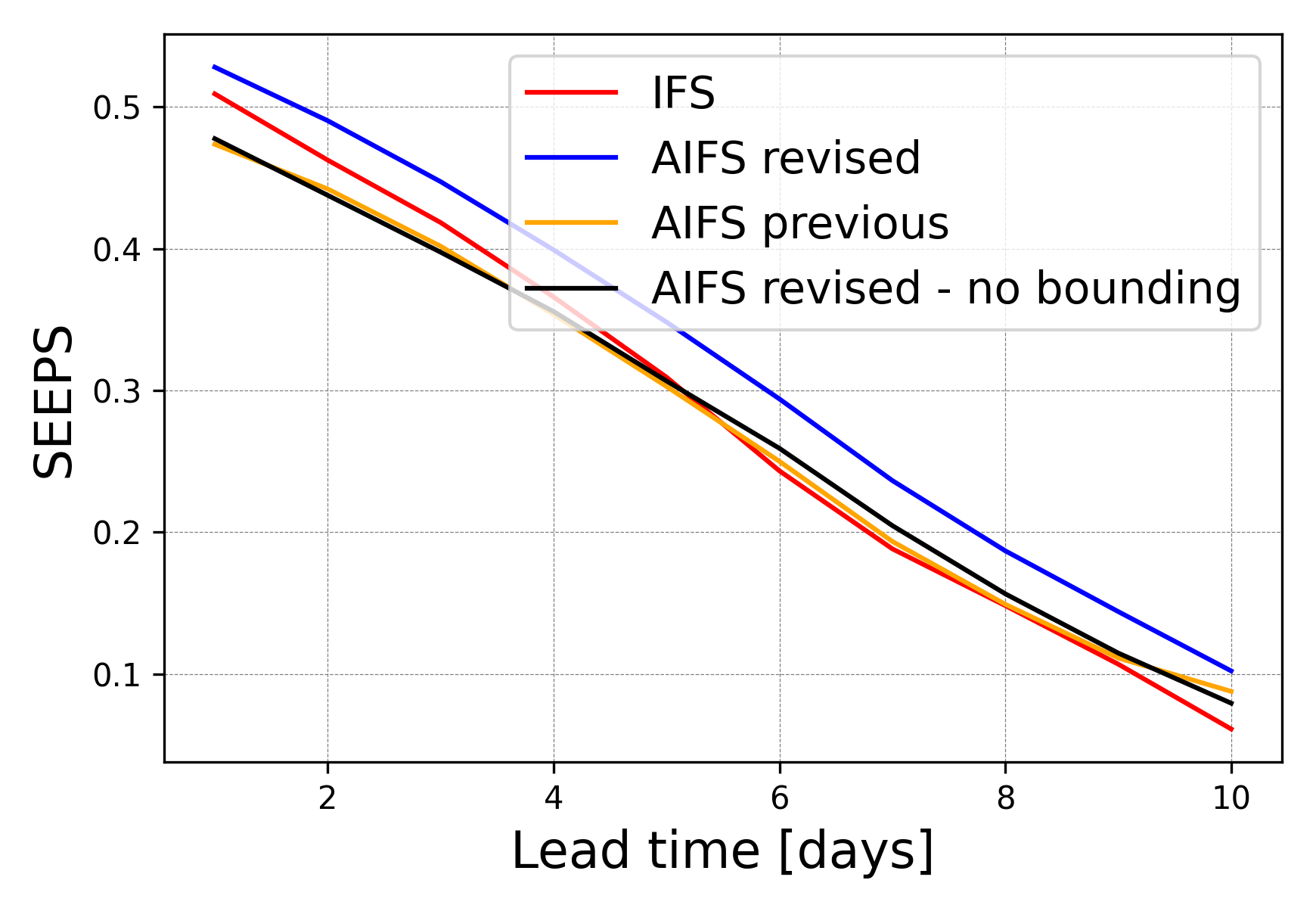}
        \caption{Northern Hemisphere }
    \end{subfigure}
    \begin{subfigure}[b]{0.495\textwidth}
        \centering
        \includegraphics[width=\linewidth, page = 5]{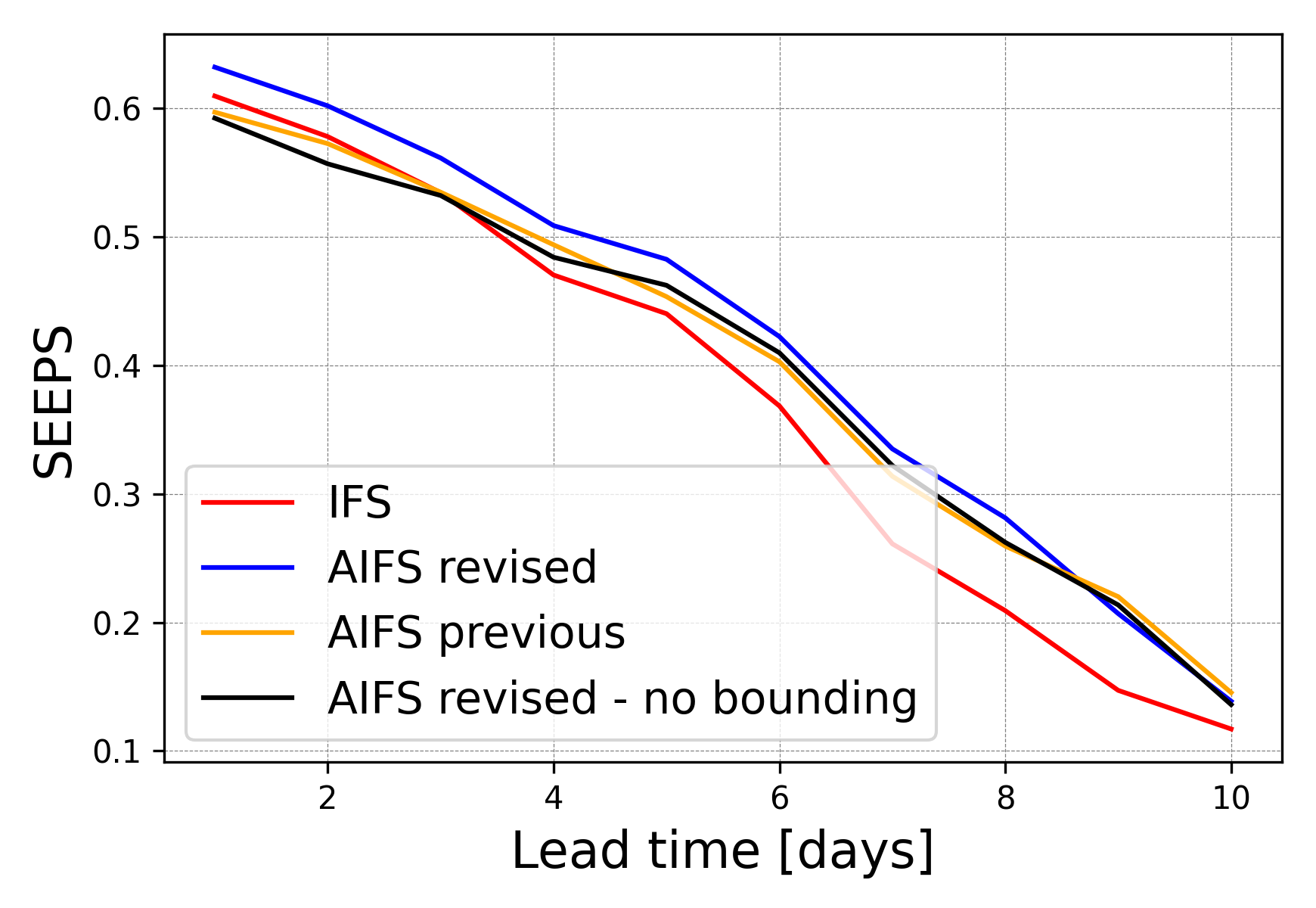}
        \caption{Southern Hemisphere}
    \end{subfigure}

    \caption{SEEPS skill scores for 2023 JJA comparing revised AIFS (blue), revised AIFS without bounding (black), previous AIFS (orange), and IFS (red) across different regions. The improvement observed in total precipitation forecast skill in the revised AIFS version can mainly be attributed to bounding the output of the model.}
    \label{fig:seeps_scores_bounding}
\end{figure}

Other factors that might improve the precipitation forecast skill in the revised AIFS version are the inclusion of additional variables, the improved learning rate scheduling for rollout fine-tuning and the expansion of the training dataset. To isolate the effect of the bounding, we retrained the revised AIFS version without bounding the total precipitation output. The SEEPS skill score for the June-July-August 2023 season is shown in Figure \ref{fig:seeps_scores_bounding}. The results show that the improvement observed in total precipitation forecast skill in the revised AIFS version can mainly be attributed to constraining the output, since the revised AIFS version without bounding performs similarly to the previous AIFS version.

\begin{figure}
    \centering
    \includegraphics[width=\linewidth]{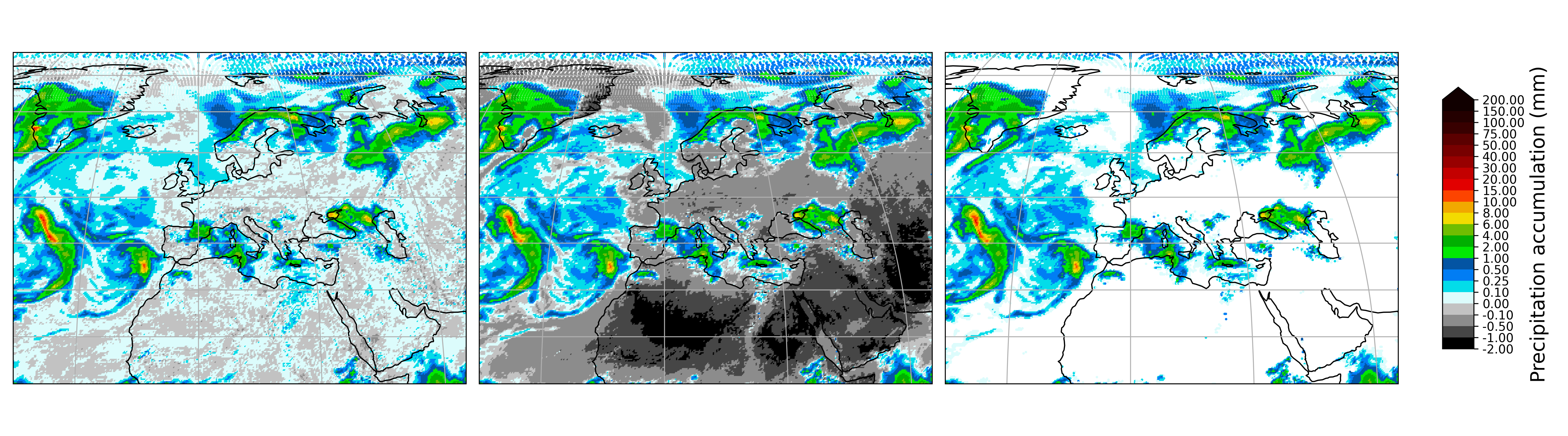}
    \vspace{0 em} 
    \begin{minipage}{\linewidth}
        \centering
        \hspace{-5em} (a) AIFS previous \hspace{3em} (b) AIFS rev. (neg. space) \hspace{3em} (c) AIFS revised 
    \end{minipage}
\caption{Comparison of 6-hour total precipitation from previous AIFS, revised AIFS without the final ReLU layer to show the negative space, and the standard revised AIFS with the final ReLU layer. Forecasts are initialised at 01/06/2023 00:00 UTC and valid at 01/06/2023 06:00 UTC. Removing the final bounding layer from the AIFS revised model reveals the behaviour of the negative space for the total precipitation variable. The model has implicitly learned to use the negative space as a proxy for ``no-rain'' classification.}
\label{fig:TP_NEGATIVE_SPACE}
\end{figure}

To better understand the mechanisms behind total precipitation forecasting in the revised AIFS version, we examine the model's behaviour in the negative forecast space, revealed by removing the final ReLU layer (Figure \ref{fig:TP_NEGATIVE_SPACE}). Bounding an output variable via ReLU has some drawbacks: the negative space is unconstrained since any changes in model behaviour in the negative space are mapped to zero before the loss is computed, which means that these points do not influence the weight update. Interestingly, this hidden negative space shows a coherent and structured pattern. Very dry regions, such as the Sahara Desert, exhibit strongly negative values, while areas near precipitation events gradually approach zero in a smooth and continuous manner. This suggests that the model has implicitly learned to use the negative space as a proxy for ``no-rain'' classification.

\begin{figure}[h]
    \centering
    \includegraphics[width=\linewidth]{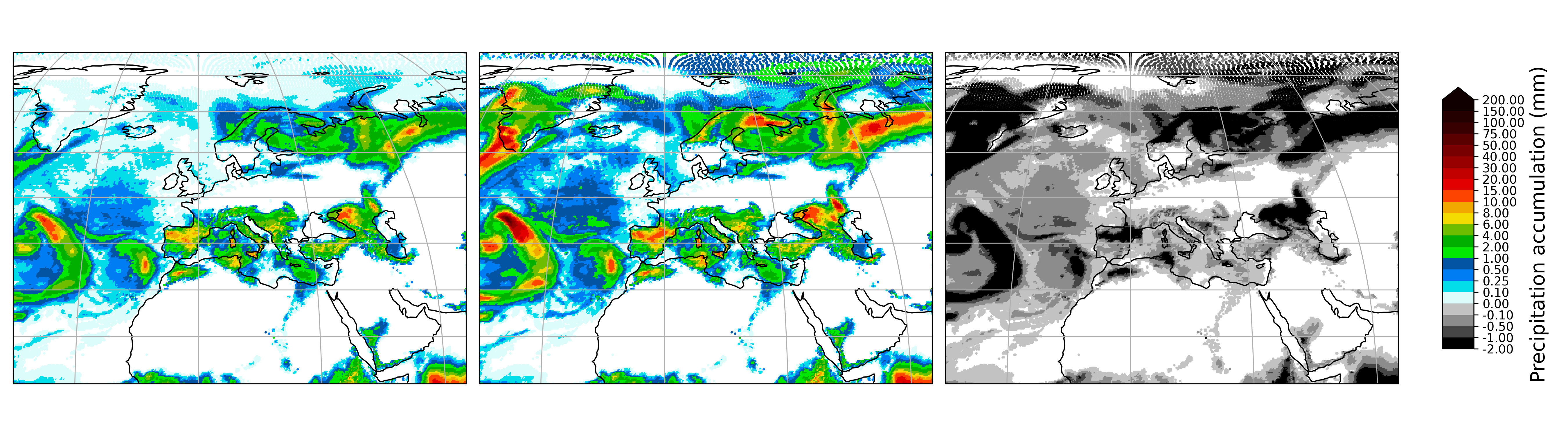}

        \vspace{0 em} 
    \begin{minipage}{\linewidth}
        \centering
        \hspace{-6em} (a) Convective precipitation \hspace{1.5em} (b) Total precipitation \hspace{2.5em} (c) Difference (cp-tp) 
    \end{minipage}
\caption{Comparison of 24-hour total and convective precipitation accumulation forecast from the revised AIFS version, together with a map showing the difference between the two of them for the forecast issued at 01/06/2023 00:00 UTC and valid at 02/06/2023 00:00 UTC. Unlike the previous AIFS version(Figure \ref{fig:TP_CP_old}), the convective precipitation forecast is now consistent with the predicted total precipitation accumulation and no coloured regions (cp$>$tp) appear in the difference plot.}
\label{fig:TP_CP_new}
\end{figure}

The physical consistency of convective precipitation forecast in respect to total precipitation can also be evaluated for a given forecast to assess the utility of the FractionBounding strategy used. Figure \ref{fig:TP_CP_new} presents the 24-hour total and convective precipitation accumulation together with a map showing the difference between the two for a forecast issued at 01/06/2023 00:00 UTC and valid at 02/06/2023 00:00 UTC. Unlike the previous AIFS version (Figure \ref{fig:TP_CP_old}), the convective precipitation forecast is now consistent with the predicted total precipitation accumulation.

\subsection{Case Studies}

Headline verification scores for the revised AIFS show significant improvements over the conventional numerical weather prediction model. However, building trust in AI forecasting requires more than strong overall metrics. Forecasters place great importance on the ability of the model to accurately and reliably predict weather phenomena. They also value physically plausible outputs and recognizable weather patterns. To support this, we show below selected case studies.

\subsubsection{Storm Éowyn}
Storm Éowyn was an unusually strong winter storm and blizzard, initially impacting much of the Gulf Coast of the United States between January 20 and January 22, 2025. This storm broke snowfall records at a number of reporting stations \citep{climate_gov_2025_snowstorm} and represented an extreme out-of-training-distribution event with no clear analogies in the ERA5 reanalysis or the IFS Operational analysis dataset.

\begin{figure}[h]
    \centering
        \includegraphics[width=0.8\linewidth]{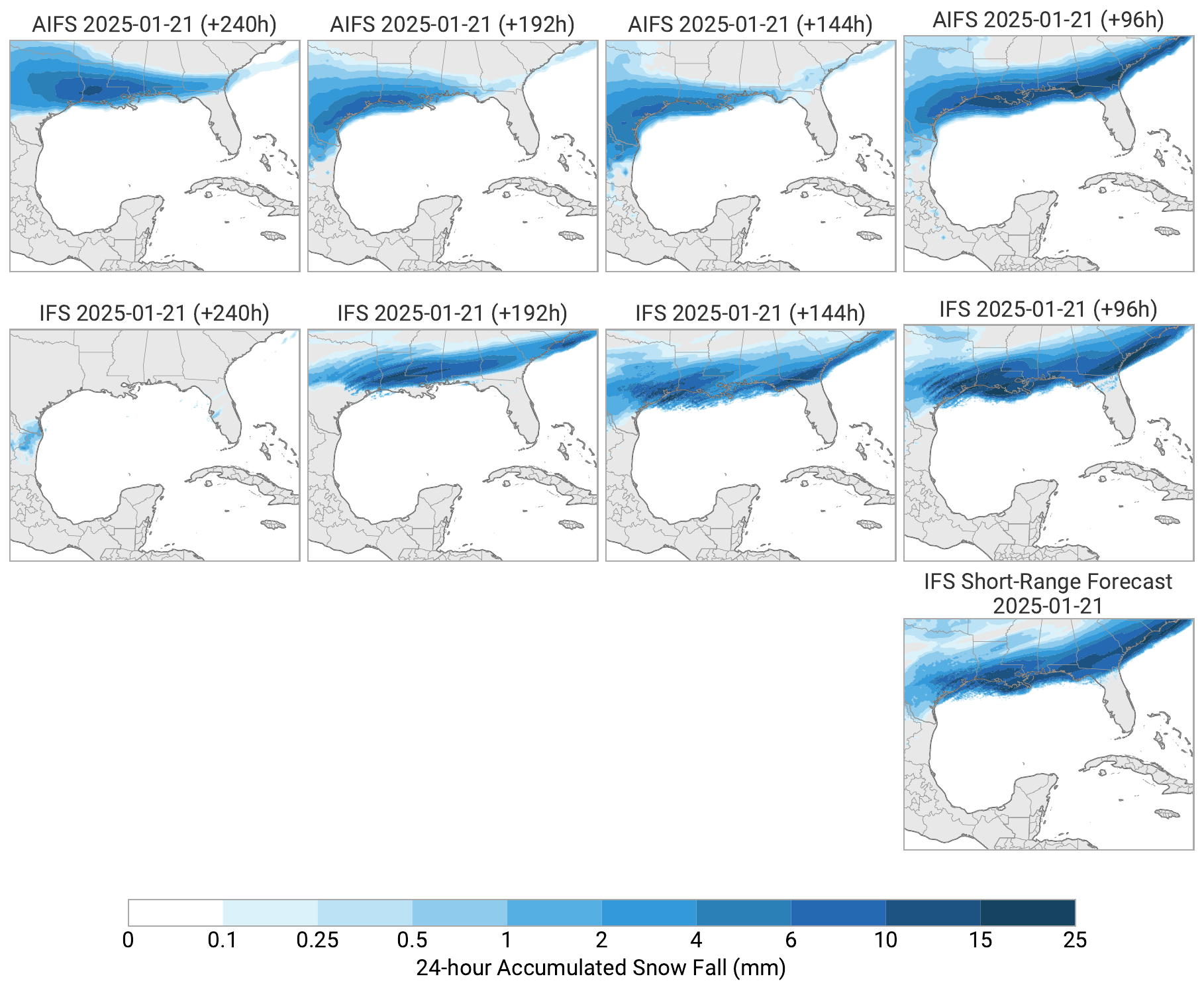}
\caption{Snowfall forecasts for AIFS (top row) and IFS (middle row) over the Gulf Coast of America at 10, 8, 6 and 4 day lead times from left to right respectively, against IFS short-range forecasts for the snowfall event (bottom row). The figure shows how the snowfall event was forecast accurately four days ahead by both the IFS and AIFS. The AIFS forecasted the event even 10 days ahead.}
\label{fig:Gulf_of_Mex_case}
\end{figure}

Figure \ref{fig:Gulf_of_Mex_case} shows the AIFS and IFS forecasts at decreasing lead times for the affected area versus the corresponding IFS short-range forecast. The AIFS delivers an accurate forecast of snowfall for this extremely rare event. This showcases the ability of the model to accurately interpret meteorological patterns and forecast physically plausible events, even if they are far from the training data. The AIFS predicted the event with a lead time of 10 days, earlier than the IFS. 

\subsubsection{Tropical Low and extreme precipitation totals in Queensland Australia}

\begin{figure}[h]
    \centering
    \includegraphics[width=1\linewidth]{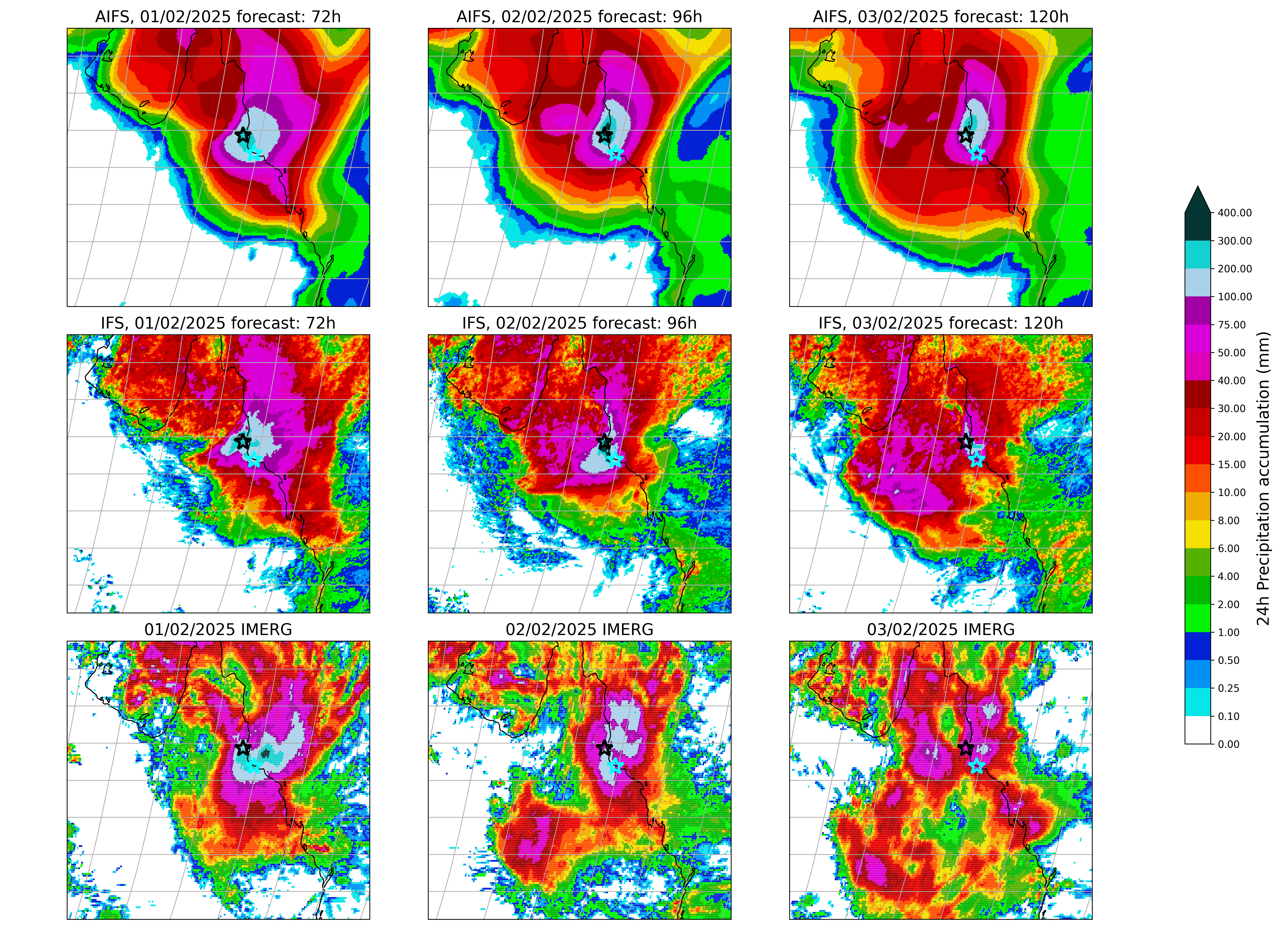}
\caption{24-hour accumulated precipitation forecasts from the AIFS (top row) and IFS (middle row) models, compared with IMERG observational data (bottom row) over northeastern Queensland for 01/02/2025 to 03/02/2025. Forecasts are initialised on 30/01/2025. The black star marks the Cardwell Range, where rainfall totals exceeded 1600~mm over the week, and the cyan star marks the city of Townsville. Both models captured the core of the extreme rainfall event, with accumulations exceeding 300~mm in 24~hours in some areas.}
\label{fig:Queensland}
\end{figure}

Starting in late January 2025, a slow-moving summer storm brought exceptional rainfall along the northeastern coast of Queensland, Australia. Within a week, rainfall accumulation totalled more than 1000 millimetres in some areas, according to the Bureau of Meteorology as reported in \cite{nasa2025queensland}. The city of Townsville saw the equivalent of six months of rain in just three days and the largest weekly rainfall total was measured at a gauge in the Cardwell Range, southwest of Tully, where nearly 1700mm fell (\cite{nasa2025queensland}, Bureau of Meteorology measurements). Figure~\ref{fig:Queensland} compares forecasts from AIFS and IFS against the IMERG \cite{IMERG} final product for the period 01/02/2025–03/02/2025. Both model forecasts were initialized on 30/01/2025, two days prior to the event. The Cardwell Range is indicated by a black star, and the city of Townsville by a cyan star. Both IFS and AIFS successfully captured the event, with 24-hour rainfall accumulations exceeding 300~mm in some regions. However, the AIFS forecast exhibits a somewhat persistent signal in the 5-day lead time, predicting very high rainfall totals near the Cardwell Range. This highlights that, despite AIFS’s tendency toward excessive spatial smoothing, it remains capable of accurately forecasting extreme events at medium range.

\section{Discussion and conclusion}\label{sec:discussion}

The revised AIFS version (1.1.0) presented here improves upon the pre-operational release through a revised training regime with more data, new forecast variables, improved stratospheric loss weights, and a bounding strategy that enforces physical constraints on the output variables. Overall, this leads to improvements of around 4–6~$\%$ across all variables, lead times, and pressure levels. The largest improvements, up to 12$\%$ gains in normalized difference in the short range,  are observed in total precipitation forecasting, which benefits from the newly introduced bounding. We showed that this has a significant impact on the prediction of no rain and light precipitation. The model displays good forecast performance for out-of-training-sample case studies, accurately capturing extreme precipitation and snowfall events.

Data plays a crucial role in the performance of AI models. Most of the improvements non-related to precipitation in the revised version of the AIFS stem from the expansion of the training dataset and the use of more recent operational ECMWF analyses for rollout fine-tuning. Since the AIFS relies on these analyses for real-time forecasting, it is important to fine-tune them regularly using up-to-date data. Regular fine-tuning with recent ECMWF analyses helps the models to adapt to shifts in the data due to new IFS model cycles.

The bounding strategy implemented also plays a crucial role, especially for precipitation forecasting. Hard-constraining model outputs increases the physical realism of forecast fields and the light-precipitation forecast skill. This improvement is attributed to a shift in the forecast space, where bounding the output facilitates the prediction of ``no-rain''. We hypothesise that the constraint enables the model to treat the negative space as likelihood of ``no-rain'' conditions, thereby facilitating the learning of the zero tp forecast. Other bounding functions may be used, and we plan to explore these in future work. In particular, we aim to investigate LeakyReLU-based approaches, which allow for weight updates with changes in the negative space, something that standard ReLU functions do not permit.

Rollout fine-tuning also emerges as a key factor shaping the forecasting skill of the model, particularly through its influence on the smoothing of outputs. While smoothing is already present in the pre-trained model, rollout fine-tuning enhances this behaviour. This reflects the model's adaptation to the inherent forecast uncertainty for longer lead times. As the model is exposed to lead times up to 72h, the minimization of the mean squared error inevitably results in an enhanced blurring of the fields. The impact of training hyperparameters on the smoothing characteristics of the output remains to be further explored. Limited testing has shown that factors such as learning rate scheduling, number of steps and the rollout strategy all have an influence on the intensity of blurring in the fields. These findings highlight an important design trade-off in training deterministic AI forecasting models: between forecast realism and optimization of MSE-based verification scores. Forecast realism refers to how physically plausible and meteorologically coherent AI-generated forecasts are, here specifically in terms of their spectral characteristics (e.g. power spectra across spatial scales). One way to evaluate the physical realism of the resulting forecasts is to assess whether their spectral signature resembles that observed in the analysis. While aggressive rollout fine-tuning strategies (such as the one used in \cite{bodnar2025foundation}) can significantly boost headline scores, here we have chosen an approach that maintains a subjective compromise between forecast realism and forecast skill measured by RMSE.

Alongside making updates to the training schedule, we have also added new variables to the AIFS while achieving improvements in forecast skill for headline atmospheric metrics. However, it remains to be seen if adding more variables and earth-system components will eventually require an increase to the latent space of the model. The additional earth-system and energy-sector variables in AIFS establish a foundation for future extensions, including ocean and wave components, expanding the number of cryospheric processes with enhanced snow modelling, and increasing the hydrological capabilities of the model. These new variables are currently taken from a consistent data source with the rest of the model variables. In the future, there is the potential to look at datasets tailored to specific earth-system components, such as ERA5-Land \citep{munoz_era5land} and the ocean and sea-ice reanalysis system (ORAS6) \citep{zuo_oras6}. AIFS currently operates at approximately 0.25° spatial resolution with a 6~hour timestep, and future work will focus on increasing both spatial and temporal resolution.

The AIFS development has now transitioned to the new Anemoi framework \citep{lang2024aifs, nipen2024regionaldatadrivenweathermodeling,wijnands2025stretchgrid}. Anemoi provides tools for the whole data-driven modelling workflow, from the generation of training datasets, to scalable probabilistic training \citep{lang2024aifs-crps} and running real-time inference with such models. Anemoi also allows for the cataloguing and archiving of model and data checkpoints to ensure reproducibility and traceability of training and inference runs and ensure that any models developed within this framework have a clear lineage. The Anemoi framework is now being used by an increasing number of Member States of ECMWF and collaborating organisations supported by ECMWF.

After a successful experimental phase, AIFS has transitioned to operational status at ECMWF on the 25th of February 2025. It is supported 24/7 alongside ECMWF's physics-based system, the IFS. The MSE trained model is labeled AIFS Single, and its forecasts are available earlier than the ones from the physics-based model chain, due to the fast runtime of AIFS. 
Results presented in this paper show that AIFS forecasts are highly skilful and they outperform the IFS forecasts across the vast majority of lead times and variables. They  highlight the relevance of AIFS for weather prediction. Future developments will focus on including more surface variables and exploring a wider range of applications such as climate reanalysis.
The operational release of the AIFS demonstrates the commitment of ECMWF to pursue the best possible weather forecasts with both physics-based and machine learning methods.

\section{Code and Model Availability}
We plan to release the configuration settings used to train this new version of AIFS Single at \url{https://github.com/ecmwf/anemoi-configs}, along with the model weights, under an open-source licence. This contribution will extend ECMWF’s collection of models available on Hugging Face at \url{https://huggingface.co/ecmwf}. The AIFS Single model data are freely available under ECMWF's Open Data Creative Commons licence (\url{https://www.ecmwf.int/en/forecasts/datasets/open-data}) and forecast charts can be seen at \url{https://charts.ecmwf.int/?query=aifs-single}.
Further details on the model’s operationalization and data dissemination can be found at \url{https://confluence.ecmwf.int/display/USS/Implementation+of+AIFS+Single+v1.0}.

\section{Acknowledgements}
We acknowledge PRACE for awarding us access to Leonardo, CINECA, Italy. We acknowledge the EuroHPC Joint Undertaking for awarding this work access to the EuroHPC supercomputer MN5, hosted by BSC in Barcelona through a EuroHPC JU Special Access call.
Ewan Pinnington's contribution is funded under the CERISE project (grant agreement No101082139), CERISE is funded by the European Union. Ana Prieto Nemesio's contribution is partially funded under the RODEO project (grant agreement: No101100651), RODEO is funded by the European Union. Views and opinions expressed are however those of the author(s) only and do not necessarily reflect those of the European Union or the Commission. Neither the European Union nor the granting authority can be held responsible for them.

\end{document}